\documentclass[pra,aps,twocolumn,groupedaddress,nofootinbib,superscriptaddress]{revtex4-2}
\usepackage{graphicx}
\usepackage[nointegrals]{wasysym} %
\usepackage[export]{adjustbox}
\usepackage{amsmath,amsfonts,amssymb,latexsym}
\usepackage{hhline}
\usepackage{bm}
\usepackage{verbatim}
\usepackage{enumitem}
\usepackage{mathrsfs}
\usepackage{slashed}
\usepackage{empheq}
\usepackage{caption,subcaption}
\usepackage{hyperref}
\usepackage{braket}
\newcommand{\bfk}{\mathbf{k}}
\newcommand{\bfq}{\mathbf{q}}
\newcommand{\hhf}{H_{\text{C}}^{\text{HF}}}
\newcommand{\bfg}{\mathbf{G}}
\newcommand{\bfG}{\mathbf{G}}
\newcommand{\bfK}{\mathbf{K}}

\newcommand{\moire}{moir\'{e} } 

\usepackage{xcolor}

\DeclareMathOperator\Log{Log}

\begin{document}
	
	\title{ Strong correlations in ABC-stacked trilayer graphene: \\ Moir\'{e} is important}
	
	\author{Adarsh S. Patri and T. Senthil}
	\affiliation{Department of Physics, Massachusetts Institute of Technology, Cambridge, MA, 02139}
	
	\begin{abstract}
	
Recent experiments on multilayer graphene materials have discovered a plethora of correlated phases, including ferromagnetism and superconductivity, in the absence of a moir\'{e} potential. These findings pose an intriguing question of whether an underlying moir\'{e} potential plays a key role in determining the phases realizable in tunable two-dimensional quantum materials, or whether it merely acts as a weak periodic potential that perturbs an underlying correlated many body state. In this work, employing a Hartree-Fock mean field analysis, we examine this question theoretically by quantitatively studying the effects of an hexagonal Boron Nitride (h-BN) substrate on ABC-stacked trilayer graphene (ABC-TLG). For the topologically trivial regime, we find that the {moir\'e} potential leads to a strong suppression of the ferromagnetism of the underlying metal. Further, band insulators appear solely at full filling of the moir\'{e} unit cell, with a moir\'{e} potential stronger than is conventionally assumed. Thus the observed correlated insulating phases in ABC-TLG aligned with h-BN cannot be understood through band folding of the ferromagnetic metal found without the moir\'{e} potential. For the topologically non-trivial regime, we discover the appearance of prominent incompressible states when fractional hole fillings (of the moir\'{e} unit cell) coincide with the occurrence of fractional-metallic states in the moir\'{e}-less setting, as well as a slight weakening of the ferromagnetic nature of the phases; however this once again requires a moir\'{e} potential stronger than is conventionally assumed. Our findings highlight the importance of interactions in renormalizing the electronic bandstructure, and emphasizes the key role played by the moir\'{e} potential in determining the strong correlation physics.

	\end{abstract}
	
	\maketitle

	\section{Introduction }
	\label{sec_intro}

The quest to develop a deeper understanding of the intrinsic nature of interacting electronic systems has undergone a resurgence in recent years, in part due to the tremendous discoveries of correlated physics in highly controllable two-dimensional materials \cite{andrei2021marvels}.
 Starting with the remarkable initial report of correlation-driven insulators and superconductivity \cite{cao_pablo_ci_tbg, cao_pablo_sc_tbg, yankowitz2019tuning, lu2019superconductors} in twisted bilayer graphene (TBG), other fascinating phenomena including ferromagnetism, electronic nematicity, fractional chern insulators, and strange metallic behavior over a wide range of tunable dopings and magnetic fields (for a sample of papers, see Ref. \cite{sharpe2019emergent, serlin2020intrinsic, saito2021hofstadter, wu2021chern, sc_nematicity_tbg, oh2021evidence, fci_tbg_2021, linear_t_2019, sm_tbg_pablo, sm_tbg_efetov}) have been found. 
 
Correlated insulators were also discovered early on in the \moire superlattice formed by ABC trilayer graphene aligned with an hexagonal Boron-Nitride substrate \cite{gate_tunable_correlated_insulator,feng_wang_abc_tlg}. This is a particularly intriguing platform where the bandwidth \cite{gate_tunable_correlated_insulator,feng_wang_abc_tlg,briding_hubbard_senthil} as well as the band topology \cite{nearly_flat_bands_senthil,moire_band_abc_tlg_theory} can be tuned by an applied perpendicular electric field. There have since been discoveries of correlated phases in a variety of related two-dimensional platforms, including {twisted double bilayer graphene \cite{tdbg_ci_2020, tdbg_sp_2020, tdbg_tuneable_ci_sp_2020, tdbg_sb_2021,tdbg_ci_2019}}, alternately twisted multilayer graphene \cite{ttg_electric_field_tunable, park2021tunable, inplane_fields, siriviboon2021abundance, kim2021spectroscopic,park2021magic, zhang2021ascendance}, and \moire transition metal dichalcogenide materials \cite{tang_simulation_2020,regan_mott_2020, xu_correlated_2020, huang_correlated_2021}.

The occurrence of correlated phenomena in a variety of such {moir\'{e}} two-dimensional systems is ultimately tied to the triumph of interaction energy scales over kinetic energy, which is enabled (for example) by the flattening of the electronic bandstructure in the \moire superlattice. 
However, an intriguing variation to this paradigm is seen in the correlated phases observed in non-{moir\'{e}} platforms.
In isolated ABC-trilayer graphene (i.e. {moir\'{e}}-less ABC-TLG, where the h-BN substrate is not aligned with ABC), correlations drive various kinds of flavor ferromagnetic metallic states \cite{tlg_abc_young}. Proximate to transitions into these phases, unconventional superconductivity has also been reported \cite{andrea_young_abc_tlg_sc}. Broadly similar phenomena are also seen in (unaligned with h-BN substrate) Bernal bilayer graphene \cite{bernal_ab_mag_young, bernal_ab_pablo, bernal_trigonal_wrap, bernal_ab_soc}. 
How does one understand the origin of the correlated phases in such non-{moir\'{e}} platforms? 
Are they related to the correlated phases seen in their {moir\'{e}} counterparts?
What role, if any, is played by the {moir\'{e}} potential in determining the low-energy phenomena? 

To provide a sharper focus to such questions, we first recapitulate the salient aspects from recent experimental studies on ABC-stacked tri-layer graphene systems, 
where the thermodynamic properties were examined by performing measurements of the electronic compressibility, $\kappa$, under the influence of (i) a perpendicular displacement field, (ii) an external magnetic field, and (iii) the {moir\'{e}} potential from an aligned hexagonal Boron Nitride (h-BN) substrate \cite{tlg_abc_young}.
At zero magnetic field and without an h-BN substrate, a sharp insulating peak was observed at CNP at a range of displacement fields, with regions of constant compressibility (on both electron and doped regimes away from CNP) separated by strongly negative $\kappa^{-1}$ {streaks that are indicative of strong interaction effects} \cite{neg_comp_2deg}. 
The nature {of these regions of constant compressibility} were discerned by applying in-plane and out-of-plane magnetic fields{, and examining the evolution of the boundaries (i.e. the negative compressibility streaks).} 
From the observed Landau level degeneracy, as well as magnetoresistance quantum oscillation frequencies, it was revealed that on the electron-doped regime, flavor-{polarized} metals ($\frac{1}{4}$-metal that is spin and valley polarized and a $\frac{1}{2}$-metal with only the spin being polarized) were being realized.
The hole-doped region of the phase diagram is significantly more complicated with a variety of phases where isospin symmetries are broken in conjunction with Lifshitz transitions of the Fermi surface topology. 
Indeed, these hole-doped phase boundaries were seen to be more strongly first-order (from marked hysteresis in Hall and resistivity measurements under applied magnetic and electric displacement fields) than the electron-doped phase boundaries.

Earlier work \cite{gate_tunable_correlated_insulator,feng_wang_abc_tlg} on ABC-TLG aligned with h-BN (which leads to a \moire superlattice) had found incompressible states appeared at $\nu = \pm 1, \pm 2$ in {moderate displacement fields (in both directions); at smaller field strengths, band insulators also appeared at $\nu  = -4$, which disappeared as the displacement field was increased, while the insulators at $\nu = -1, -2$ remained.
Here, $\nu = n_e / n_s$, is the hole filling per \moire unit cell (with $n_s$ being the density corresponding to one electron per spin-valley flavor per \moire unit cell), with $\nu = \pm 4$ referring to full filling.
{Moreover the $\nu =-1$ insulator was found \cite{feng_wang_abc_tlg} to be a Chern insulator with Chern number $|C| = 2$ for one sign of a perpendicular displacement field.}}
{For the other sign of the displacement field, the correlated nature of the $\nu = -2$ insulator was also demonstrated in a recent spectroscopic study, where the structured signature of the optical photocurrent provided strong indication of its identity as a strongly correlated (and not as a mere band) insulator \cite{long_ju_abc_tlg}.}

{
The appearance of insulating peaks in the presence of the {moir\'{e}} potential in Ref. \cite{tlg_abc_young}} {at $\nu = -1, -2$ coincided with the occurrence of $\frac{1}{4}$ and $\frac{1}{2}$ metals at those densities (in the {moir\'{e}}-less setting). This lead Ref. \cite{tlg_abc_young} to suggest that the {moir\'{e}} potential acts as a weak periodic potential (to merely fold the electronic bandstructure into the {moir\'{e}} Brillouin zone) and as such to not significantly alter the correlated ({itinerant flavor-polarized}) physics already present in the parent, non-{moir\'{e}}, trilayer}.
The implication of this idea is dramatic as it suggests that a plethora of correlated phases may be realized in gate-tunable multi-layer graphene systems, without the {moir\'{e}} superlattice that has been so fruitful in the aforementioned platforms.

In this work, we target this specific line of inquiry by quantitatively analyzing the effects of the {moir\'{e}} potential on ABC-stacked trilayer graphene systems.
{Though recent theoretical works have focused on the various broken symmetry phases and its relation to the observed superconductivity \cite{abc_tlg_gapped_gapless_macdonal_2013, zalatel_abc_tlg, macdonald_tlg_2022, sc_abc_nm_sds, abc_tlg_sc_berg, abc_tlg_ashvin, abc_tlg_sc_zhiyu, roy_abc_sc_2022, functional_rg_macdonald_2022}, a deeper analysis with regard to the influence of the {moir\'{e}} potential has yet to be performed.}
{Motivated by Ref. \cite{tlg_abc_young} to take the approach of itinerant flavor-polarization,} we first discuss the Hartree-Fock mean field phase diagram on both the electron and hole-doped sides (about the CNP) on ABC-TLG.
We then introduce the influence of an aligned h-BN substrate that creates a {moir\'{e}} potential for both signs of a perpendicular displacement field.
Our quantitative analysis provides insight into the effects of the {moir\'{e}} potential.

Earlier works \cite{nearly_flat_bands_senthil,briding_hubbard_senthil,feng_wang_abc_tlg,qimiaio_abc_hBN_2020,Pantale_n_2021,Bascones_2022} considering the interplay of electron-electron interactions with the \moire potential, focused on the effects of Coulombic interactions projected on top of the \moire minibands.
In this work, however we examine the suggestion of Ref. \cite{tlg_abc_young}, that the effects of electron-electron interactions should be considered \textit{before} examining the effects of the \moire potential. Conceptually this involves treating the effects of the electron-electron interactions as energetically more dominant than those effects from the \moire potential. Specifically this means that in the \moire system, the Coulomb interaction may strongly mix the active bands near the chemical potential with the remote bands, and thus it may be inappropriate to model the system by projecting the Coulomb interaction to just the active band. 

{Within the context of band theory \cite{ nearly_flat_bands_senthil,moire_band_abc_tlg_theory}, depending on the sign/direction of the perpendicular displacement field, the subsequent active conduction and valence bands acquire a non-zero Chern number. 
In particular, for h-BN aligned with the top layer, $D>0$ realizes an active valence band of Chern number $|C| = 3$ (the two valleys have equal and opposite Chern number), while the $D<0$ realizes a topologically trivial active valence band ($C=0$). (Allowing for a Hartree-Fock renormalization of the dispersion and Bloch wavefunctions \cite{feng_wang_abc_tlg} of the active band -- through interactions with the remote occupied bands -- allows for the valley Chern number to also take the value $|C| = 2$). 

For the topologically trivial regime, we find that the spin and valley characters of the phases undergo a strong suppression under an increasing \moire potential. 
This suppression leads to the appearance of incompressible states solely at complete filling of the spin-valley flavors ($\nu = -4$), which is in agreement with conventional band theory expectations.
However, the insulating gap develops above a critical \moire potential $\approx 2-3$ times the expected \textit{ab initio} {moir\'{e}} potential strengths.
More importantly, the lack of insulating states at $\nu = -1, -2$ is in contrast with the experimental observations. {The implication is that the correlated insulators observed at $\nu = -2$ or $\nu = -1$ cannot be understood simply as due to the \moire potential opening band gaps at the chemical potential of a flavor polarized metal. Thus the opposite limit of treating these insulators as arising from strong correlations on the active \moire bands may be more appropriate, as assumed in the early literature. Indeed, recent spectroscopic results \cite{long_ju_abc_tlg} support such a picture.
}

For the topologically non-trivial regime, we find that insulating states appear at fractional fillings of the \moire unit cell depending on the strength of the displacement fields.
For small displacement fields, insulators appear at one-quarter ($\nu = -1$), two-quarters ($\nu = -2$), three-quarters ($\nu = -3$) and at full filling ($\nu = -4$).
With increasing displacement fields, insulators appear at fewer fractional densities: insulators at $\nu = -1, -2, -3$; to insulators only at $\nu = -1, -2$; to insulators solely at full-filling $\nu = -4$.
The occurrence of these insulators coincides with having $\frac{m}{4}$ metallic phases at $\nu = -m$ densities in the {moir\'{e}}-less setting, a conclusion that is in agreement with the experimental findings.
However, the insulating gaps increase continuously as a function of {moir\'{e}} potential strength above a critical \moire potential strengths $\approx 4-7$ times the expected \textit{ab initio} {moir\'{e}} potential strengths.
{This required amplification of the \moire potential (as compared to \textit{ab initio} estimates) to realize the insulators indicates the presence of strong correlation effects (beyond that captured from mean-field) even in the topologically non-trivial regime.}
{Finally, we also find that the hole-valence band associated with the $\nu = -1$ insulator carries a Chern number of $|C| = 3$.

The evidently asymmetrical behaviors for the topologically trivial and non-trivial regimes reinforces our emphasis on the importance of \moire potentials in understanding the correlated phases in this system, and hence more generally in two-dimensional \moire materials.

}

{
The remainder of the manuscript is organized as follows.
In Sec. \ref{sec_model}, we present the model and energetic dispersions of ABC-TLG, emphasizing the enhanced density of states due to the flat band edges.
We next discuss the Hartree-Fock decoupling and the variety of ansatzes that are considered in Sec. \ref{sec_hf_decoupling}.
In Sec. \ref{sec_hf_phase_diagram}, we analyze the Hartree-Fock phase diagram on both the electron and doped regimes about the CNP.
In Sec. \ref{sec_moire}, we introduce the {moir\'{e}} potential and demonstrate the development of the insulating states as well as the modification to the Hartree-Fock phase diagram for both the topologically trivial and non-trivial regimes.
Finally, we conclude and propose directions of future work in Sec. \ref{sec_discussion}.

\section{Model of ABC-stacked TLG}
\label{sec_model}

	\begin{figure}[t]
	\centering
		\includegraphics[width=1\linewidth]{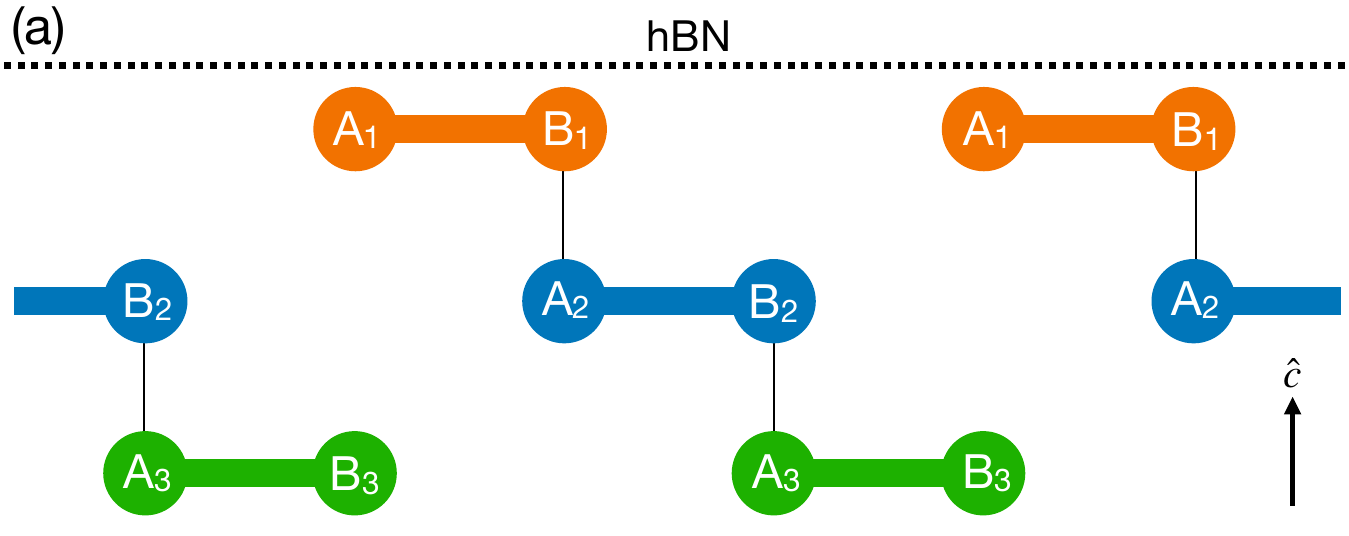} \hspace{5mm} \\
		\includegraphics[width=0.95\linewidth]{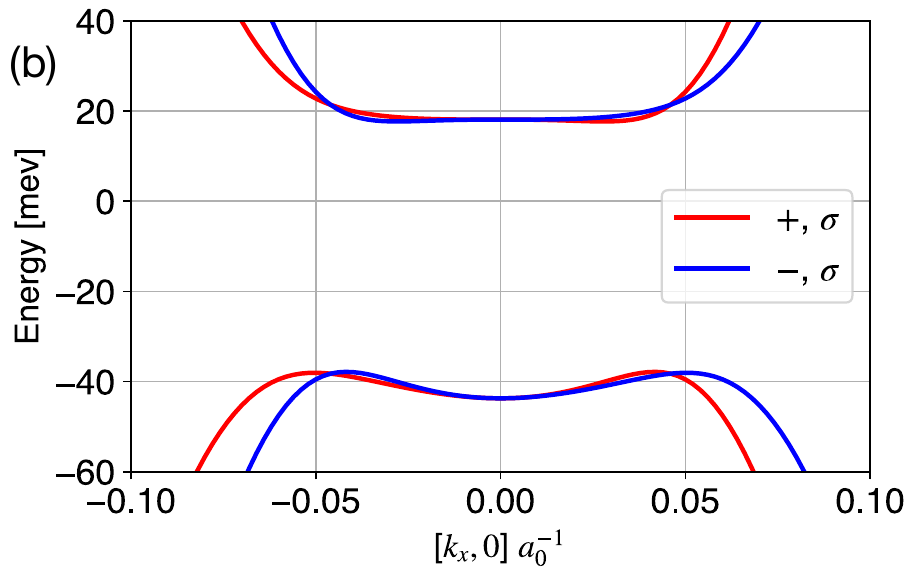}
		\caption{ABC-TLG geometry and non-interacting bandstructure.
		(a): ABC-TLG aligned with h-BN. The solid lines indicate the large dimerization between A and B sublattices. 
		One unit cell includes A and B sublattices from each of the three layers.
		The h-BN layer is aligned with the top layer in this work.
		(b): Non- interacting ABC-TLG band structure under a perpendicular displacement field ($u_d = 30$meV) within the UV cutoff $0.1 a_0^{-1}$ for valley $\tau = \{ +, - \}$ and spin $\sigma=\{\uparrow, \downarrow \}$.
		The flat band edges lead to Van-Hove singularities in the density of states.}
		\label{fig_non_int_band_structure}		
	\end{figure} 
In ABC-stacked trilayer graphene, three honeycomb layers (i.e. graphene monolayers) are in-equivalently aligned on top of each other in the $c$-axis direction.
The inequivalent stacking is achieved by shifting each layer (relative to the layer beneath) by a constant translation vector.
Each unit cell of ABC-TLG consists of six sublattice sites ($A$ and $B$ from each of the three layers) and are indexed as $A_i$ or $B_i$, where $i \in [1,3]$ denotes the layer, {as depicted in Fig. \ref{fig_non_int_band_structure}(a)}.

In monolayer graphene, one recalls that the low energy electrons possess possess valley flavor ($K$ and $K'$ henceforth denoted as $\pm$) as well as spin ($\uparrow$ and $\downarrow$).
Combining sublattice-layer ($\times 3 \times 2$), valley ($\times 2$), and spin ($\times 2$) degrees of freedom, the low-energy electrons in ABC-TLG possess in-principle $24$ flavor degrees of freedom in total.
Fortunately, the examined {active degrees of freedom} can be significantly reduced due to a large dimerization potential between $B_1$ and $A_2$ and $B_2$ and $A_3$ sites \cite{macdonald_abc_tlg_band_structure, abc_stack_projection_2016}.
As such, the low-energy electronic degrees of freedom are confined to reside on the $A_1$ and $B_3$ layer sites, effectively reducing the dimension of the electronic Hilbert space to $2 \times 2 \times 2 = 8$, where each two-dimensional subspace is formed by the layer $\sigma = \{A_1, B_3\}$, valley $ \tau = \{+, -\}$ and spin $ s = \{ \uparrow , \downarrow \}$.
The low-energy Hamiltonian in this basis is \cite{macdonald_abc_tlg_band_structure, abc_stack_projection_2016, abc_tlg_ashvin, zalatel_abc_tlg},
\begin{align}
\label{eq_h0}
H_0 = \sum_{\bfk} \sum_{s= \uparrow, \downarrow} \sum_{\tau = \pm} c_{\bfk s \tau}^{\dag} \mathcal{H}_0 ^{\tau} (\bfk) c_{\bfk s \tau},
\end{align}
where $c_{\bfk s, \tau} = (c_{\bfk, A_1,\tau,s}, c_{\bfk, B_3,\tau,s})^\intercal$, and the kernel is $\mathcal{H}_0 ^\tau (\bfk) = (\epsilon_\bfk^s - \mu_c) \sigma^0 + (\tau \alpha_\bfk ^{\text{ch}} + \epsilon_\bfk^{\text{tr}}) \sigma^1 + \beta_{\bfk}^{\text{ch}} \sigma^2 + \epsilon_{\bfk}^{\text{gap}} \sigma^3$.
Here the Pauli matrices $\sigma^j$ are in the layer space, $\mu_c$ is the chemical potential, $\alpha_\bfk ^{\text{ch}} = \text{Re}\Big[\frac{v_0^3}{\gamma_1^2}(k_x +i k_y)^3 \Big]$, $\beta_\bfk ^{\text{ch}} = \text{Im}\Big[\frac{v_0^3}{\gamma_1^2}(k_x +i k_y)^3 \Big]$, $\epsilon_{\bfk}^s = (\delta + \frac{u_a}{3}) - (\frac{2 v_0 v_4}{\gamma_1} + \frac{u_av_0 ^2}{\gamma_1^2})\bfk^2$, $\epsilon_{\bfk}^{\text{tr}} = \frac{\gamma_2}{2} - \frac{2v_0 v_3}{\gamma_1}\bfk^2 $, $\epsilon_{\bfk}^{\text{gap}} = u_d (1 - \frac{v_0^2} {\gamma_1^2}\bfk^2)$.
The displacement potential $u_d$ is set by the externally applied perpendicular electric field; we list the values of the remaining chosen parameters in Appendix \ref{app_params}.
{For brevity, we henceforth refer to $u_d$ as the displacement field (keeping in mind that it is ultimately a potential energy difference).}
We note that the non-interacting Hamiltonian is diagonal in spin and valley degrees of freedom.
In Fig. \ref{fig_non_int_band_structure}(b) we depict the associated non-interacting band structure.
Due to time-reversal symmetry, the dispersions $\xi_{\pm}(\bfk)$ of the two valleys are related by $\bfk \rightarrow -\bfk$ i.e. $\xi_+(\bfk) = \xi_-(-\bfk)$.
An important feature of the bandstructure is the existence of flat band edges in the conduction and valence bands that get enhanced under the application of a perpendicular displacement field.
The associated van-Hove singularities play a fundamental role in the interaction-driven Stoner instabilities discussed in the next section.

The low-energy electrons interact via a repulsive dual gate-screened Coulomb interaction \cite{tbg_theory_prx, tlg_abc_young},
\begin{align}
\label{eq_screened_coulomb}
H_{\text{C}} = \frac{1}{2 A} \sum_{\bfk, \bfk', \bfq} \sum_{\mu, \nu} V^{\text{sc}}_C (\bfq) c^{\dag}_{\bfk + \bfq; \mu} c^{\dag}_{\bfk' - \bfq; \nu}  c_{\bfk' ; \nu}  c_{\bfk; \mu} , 
\end{align}
where $A$ is the area of the system, $\mu, \nu$ denotes a compact layer-valley-flavor index, $V^{\text{sc}}_C (\bfq)= \frac{e^2}{2 \epsilon_0 \epsilon q} \tanh(q d_s)$ is the aforementioned screened potential (in SI units), $\epsilon$ the effective dielectric constant, and $d_s$ is the distance between the double screening metallic gates and the respective top and bottom layer.
We note that the long-wavelength $\bfq =0$ potential is given by its smooth limiting case $\bfq \rightarrow 0$.

{We also note the sole inclusion of the SU(4) symmetric (in spin-valley space) Coulombic interaction in Eq. \ref{eq_screened_coulomb}.}
{This symmetry is, of course, broken by the valley-dependent kinetic term in $H_0$ down to $\mathrm{U(2)_+} \times \mathrm{U(2)_-}$, which allows independent spin-rotation in each valley.
In addition, due to the lack of spin-orbit coupling, there is an additional spin-less time reversal that allows interchanging between the different valleys $\mathbb{T} = \tau^x K$ \cite{zalatel_abc_tlg, abc_tlg_ashvin}.
It is the combination of the time-reversal induced valley swap and the valley-independent spin rotation that allows a variety of variety of degenerate ground states of the system (as discussed in the next section).}

\textcolor{black}{The inclusion of additional, sub-dominant terms, can, however, favor some of the degenerate ground states over others.
The magnitude of these additional interactions, $J$, can be estimated \cite{tlg_abc_young} by taking them as arising from Coulombic interactions on the lattice length scale $a_0$ i.e. $J \sim \int_0^{a_0} \,d^2r \frac{e^2}{\epsilon r} = \frac{2 \pi e^2 a_0}{\epsilon} \sim \frac{a_0}{d_s} V^{\text{sc}}_C (0)$.
Since the spacing of the metallic gates is taken as $d_s = 150 a_0$, these additional interactions can be around two orders of magnitude smaller than the screened interaction.}

An important such sub-dominant term is a spin Hund's couplings of the form $\sim -\tilde{J}_H \bf{s}_{+} \cdot \bf{s}_{-}$ and $\sim -J_H \bf{s}_{+-} \cdot \bf{s}_{-+}$, where $\bf{s}_{\tau}$ is the spin density in valley $\tau$, while $\bf{s}_{\tau \tau'} \sim c^{\dag}_{\tau, \alpha} \bf{s}_{\alpha \beta} c_{\tau', \beta}$ is the inter-valley spin density (we refer the reader to Ref. \cite{zalatel_abc_tlg} for complete details on the form of the Hund's couplings).
One can see that antiferromagnetic $\tilde{J}_H<0$ favors opposite spin polarization in the two valleys, whereas ferromagnetic $\tilde{J}_H>0$ favors spins in each valley to be polarized ferromagnetically (and leaving the valley unpolarized); similarly, in an IVC state, $J_H >0$ ($J_H <0$) favors a spin polarized triplet (un-polarized singlet) state.
Though we do not explicitly include such a symmetry breaking term in this work, we can nonetheless examine its impact by examining the particular ground state (from the obtained degenerate set) whose energy is lowered by satisfying the Hund's coupling. 
As will be seen in Sec. \ref{sec_moire}, retaining the full-unbroken symmetry will be helpful in understanding when insulating states may appear under a periodic \moire potential.}

\section{Hartree-Fock decoupling}
\label{sec_hf_decoupling}

The interplay of the flat conduction and valence band edges with the screened Coulombic interaction has the potential to give rise to a variety of broken symmetry phases. In the absence of the \moire potential, the nature of the broken symmetry phases was examined within a Hartree-Fock treatment in Ref. \cite{zalatel_abc_tlg}, and shown to be in general agreement with experiment. Here we first briefly review the results of this Hartree-Fock calculation before moving to our main interest, namely the modifications induced by the \moire potential.

To examine the interaction-driven magnetic phases, the screened Coulomb interaction in Eq. \ref{eq_screened_coulomb} is decoupled into the Hartree and Fock channels,
\begin{align}
\label{eq_hf_decoupled}
H_{\text{C}}^{\text{HF}} & =  \sum_{\bfk} \sum_{\mu} \left[ \frac{1}{A} \sum_{\bfk'; \nu} \chi^{\nu \nu} (\bfk') V^{\text{sc}}_C (\bfq =0) \right] c_{\bfk; \mu} ^\dag c_{\bfk; \mu} \nonumber \\
& - \sum_{\bfk} \sum_{\mu, \nu} \left[ \frac{1}{A} \sum_{\bfk'} V^{\text{sc}}_C (\bfk - \bfk') \chi^{\nu \mu} (\bfk') \right] c_{\bfk; \mu} ^\dag c_{\bfk; \nu}.
\end{align}
The first term is the Hartree term that acts to renormalize the chemical potential in Eq. \ref{eq_h0}, while the second is the Fock term that permits mixing between the various layer-valley-spin flavors.
The above Hartree-Fock decoupling is performed assuming an ansatz of the form, $\langle c_{\bfk'; \nu}^\dag c_{\bfk; \mu} \rangle = \chi^{\nu \mu} (\bfk) \delta_{\bfk', \bfk}$, where the expectation value is taken with respect to the Hartree-Fock Hamiltonian $H_0 + \hhf$.
Indeed, as will be seen, such an ansatz allows a number of broken-symmetry phases to be explored.
In light of the lack of appreciable spin-orbit coupling in ABC-TLG, the spin degrees of freedom are decoupled from the rest and as such the ansatz is assumed to be diagonal in spin. 
The associated mean-field parameters $\chi^{\nu \mu} (\bfk)$ are solved self-consistently at a constant electron density ($n_{\text{den}}$), which is fixed at each self-consistent iteration by solving for the chemical potential i.e. $\frac{1}{A} \sum_{\bfk} \sum_{\nu} n_F(\xi^{\nu}_{\bfk}) = n_{\text{den}}$, where $\xi^{\nu}_{\bfk}$ are the associated energy eigenvalues of the Hartree-Fock Hamiltonian.
We have dropped the constant (fermionic operator-independent) terms from the decoupling in Eq. \ref{eq_hf_decoupled}.
We also note that in Hartree-Fock studies of twisted-bilayer graphene (TBG) one needs to carefully account for the logarithmic renormalization (arising from the Fock term) of the Dirac velocity near the Dirac points in the mini-bands by subtracting out the contribution from a reference state \cite{tbg_theory_prx}.
This subtlety is not of importance in ABC-TLG as the low-energy excitations, in the presence of the interlayer couplings and applied displacement field, are electrons about a Fermi surface, unlike the aforementioned Dirac nodes in TBG.

\section{Electron and Hole doped Hartree-Fock Phase diagram for $u_d >0$}
\label{sec_hf_phase_diagram}

	\begin{figure}
		\includegraphics[width=\linewidth]{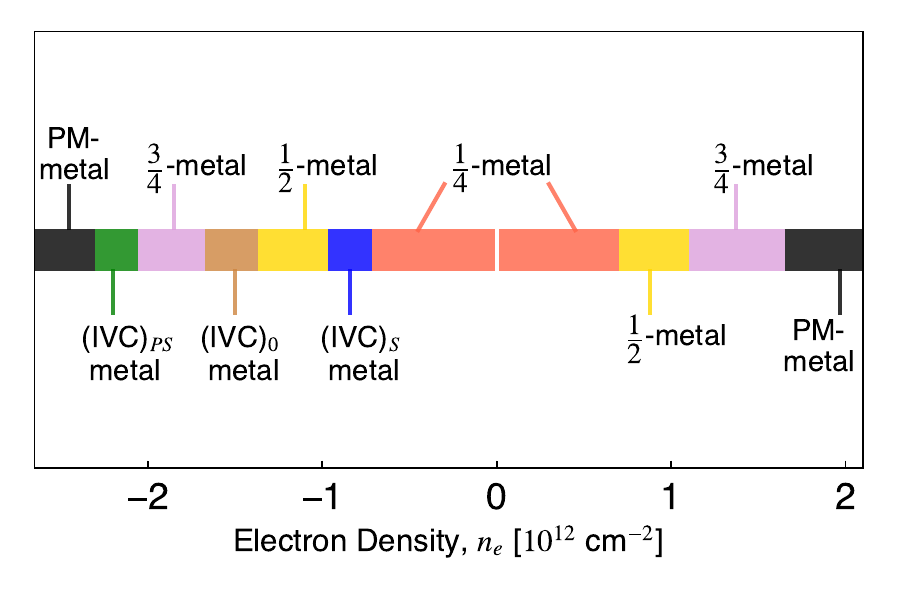}
		\caption{Electron and hole doped Hartree-Fock phase diagram for displacement field $u_d = 30$ meV.
		The phases are labelled by their spin-valley occupation number (the layer degree of freedom places the role of band index that the electrons occupy).
		The depicted phases are $\frac{m}{4}$ metal, where $m$ of four spin-valley-flavors are occupied, IVC metal where the fermions carry a linear combination of the valley degrees of freedom, and a paramagnetic metal (PM) where all four flavors are occupied. 
		The subscript for the IVC metals indicate the spin polarization: $S$ = fully spin polarized, $0$ = spin unpolarized, $PS$ = partially spin polarized (ratio of approximately two-thirds occupation for one spin quantum number and one-third for other spin quantum number).
		The CNP (indicated by white vertical line) retains the gapped insulating nature of the non-interacting model in Fig. \ref{fig_non_int_band_structure}(b).}
		\label{fig_hf_phase_diagram}				
	\end{figure} 

Equipped with the Hartree-Fock (HF) decoupling, one can capture the variety of possible broken symmetry phases as a function of the electron and hole-doping from CNP.
The doping densities, despite being remarkably small so as to begin to fill the edges of the conduction ($n_e >0$) or valence ($n_e <0$) band, can trigger a number of Stoner-like instabilities. 
We present in Fig. \ref{fig_hf_phase_diagram} the HF phase diagram for displacement potential difference of $u_d = 30$ meV; {the general features of the phase diagram carry through for varied displacement field strengths as depicted in Appendix \ref{app_u_d_positive_other}.} 
As seen, there are a number of possible phases as the electron density is tuned into the electron-doped ($n_e >0$) and hole-doped ($n_e<0$) regimes from the insulating CNP.
Indeed the nature of the broken-symmetry phases can be intuitively understood from the usual criterion of Stoner instability i.e. magnetism prevails when $U g(E_F) >1$, where $U$ is the strength of the interaction and $g(E_F)$ is the density of states at the Fermi level.
For a fixed interaction strength, and by tuning the electron density so as to occupy regions of the flat band-edges (in Fig. \ref{fig_non_int_band_structure}(b)), the associated Van-Hove singularities in the density of states permits the Stoner criterion to be satisfied, thus leading to the broken symmetry phases.
The ultimate phase realized at particular dopings is thus determined by the victor of the kinetic and Coulombic energy battle.
Please see Fig. \ref{fig_hf_phase_diagram} describing the nature of the phases and their evolution.

In the electron-doped side metallic states with uniform polarization of conserved spin and valley quantum numbers ({\em i.e}, flavor-polarized metals) are found, with (at the level of HF) first-order transitions between various $\frac{m}{4}$-metals, $m \in [1,4]$. In each $\frac{m}{4}$-metal, $m$ of the four possible flavors are occupied. In the absence of the Hund's interaction, there is a degenerate manifold of ground states for such flavor- ferromagnetic metals related by the $U(2)_+ \times U(2)_-$ symmetry. This degeneracy will be partly lifted by the weak inter-valley Hund's coupling. 

{The hole-doped regime offers a greater variety of broken-symmetry phases than the electron-doped regime.
This difference is anticipated by considering the non-interacting dispersion in Fig. \ref{fig_non_int_band_structure}(b), where the valence bands possess more structure in momentum space -- with both local minima and maxima in the dispersion -- leading to a more structured density of states.}
In addition to the $\frac{m}{4}$-metallic phases, a number of inter-valley coherent (IVC) metals are also realized.
In this context, an IVC phase is characterized as possessing electrons in a superposition of both valley degrees of freedom, with the spin degree of freedom being either fully-polarized, partly-polarized, or unpolarized.
{The valley polarization and inter-valley mixing can be characterized by $I_z$ and $I_{\perp}$ (in arbitrary units), respectively (discussed further in Appendix \ref{app_ivc_order}),
\begin{align}
I_a = \left| \sum_{\alpha, \beta, s} \sum_{\Lambda, \Omega} \frac{1}{A} \sum_{\bfq} \Big[ \langle c^{\dag}_{\Lambda, s, \alpha} (\bfq) c_{\Omega, s, \beta} (\bfq)  \rangle \tau^a _{\Lambda, \Omega} \Big] \right|,
\end{align}
where $a = \{x,y,z\}$, $\tau^a$ specifies a Pauli matrix in valley space (with $\Lambda, \Omega$ indicating the valley labels $\bfK$, $\bfK'$), $\{\alpha, \beta\}$ are over the band degrees of freedom (including the layer label), and $s$ being the spin quantum number.
Since $I_{x,y}$ are rotated into each other by the valley $U(1)$ symmetry, we define a perpendicular component to indicate the magnitude of the IVC order parameter: $I_{\perp} = \sqrt{I_x^2 + I_y^2}$.}
\textcolor{black}{We include a unit of area normalization of $[10^{16} m^2]^{-1}$ for the valley and IVC order parameters.}
Though the IVC phases may involve two or higher number of electron flavors, we choose not to label these phases in the same terminology of the $\frac{m}{4}$-metallic phases due to the lack of valley flavor conservation.

A detailed tour of the Hartree-Fock phase diagram is reviewed in Appendix \ref{app_hf_no_moire}.

\section{Influence of {moir\'{e}} potential}
\label{sec_moire}

 We are now ready to turn our attention to the effects of the {moir\'{e}} potential by aligning ABC-TLG with an underlying hexagonal Boron Nitride (h-BN) substrate. 
 A {moir\'{e}}-superlattice (with its associated superlattice potential) is then generated due to the lattice mismatch between graphene and h-BN. 
This periodic potential reconstructs the original HF band structure by folding the original bands into mini-bands within the mini-Brillouin zone (mBZ).
For instance, aligning two layers with different lattice constants, $a_{1,2}$, generates a {moir\'{e}}-pattern with lattice constant $a_m = \frac{a}{\xi}$, where $\xi = \frac{a_1 - a_2}{a_2}$ is the normalized-difference in layer lattice constants.
For the purpose of this work, we examine the situation where the subsequent {moir\'{e}} lattice constant is $a_m = 60 a_0 \approx 15 \text{nm}$, as appropriate to ABC/h-BN.

The {moir\'{e}}-superlattice potential acts primarily on the ABC-TLG layer in immediate contact with the h-BN.
{We consider two scenarios for the h-BN alignment with the top-layer (i) with $u_d >0$, and (ii) with $u_d<0$.
The non-interacting ABC-TLG Hamiltonian (Eq. \ref{eq_h0}) under the presence of a top-layer aligned \moire potential realizes a topologically non-trivial phase for $u_d >0$, and a trivial phase for $u_d <0$ \cite{nearly_flat_bands_senthil}; accounting for electron-electron interactions on top of the \moire bandstructure leads to preserving the non-trivial topology of the valence band (albeit with a reduced Chern number \cite{feng_wang_abc_tlg}).
We contrast this with our study where the effects of electron-electron interactions are accounted for \textit{before} the implementation of the \moire potential.
Nonetheless, for brevity, we label the two scenarios $u_d <0$ ($u_d>0$) as topologically (non-) trivial. }

We recall that the large dimerization potential between $B_1$ and $A_2$ sites in ABC-TLG led to a low-energy effective description in terms of layer $A_1$ and $B_3$ sites.
As such, the projected {moir\'{e}}-potential is of the form \cite{moire_band_abc_tlg_theory_macdonald, nearly_flat_bands_senthil, briding_hubbard_senthil},
\begin{align}
\label{eq_moire_potential}
H_{m} = V_m \sum_{\tau, s} \sum_{\bfk, \bfg} \left[ H_0 (\bfg) + H_z(\bfg) \right] c_{\bfk + \bfg, {A_1} , \tau, s}^{\dag} c_{\bfk, {A_1}, \tau, s}. 
\end{align}
Here $V_m$ is a dimensionless number denoting the strength of the \moire potential relative to that estimated from \textit{ab initio} calculations. (Thus $V_m = 1$ corresponds to the \textit{ab initio} \moire potential term). \textcolor{black}{$ \bfk  = \bfk + \bfg$ are momentum vectors}, $\bfg = m \bfg_+  + n \bfg_-$ are {moir\'{e}}-reciprocal lattice vectors spanned by reciprocal basis vectors $\bfg_{\pm} = \frac{4 \pi}{\sqrt{3} a_m} (-\frac{\sqrt{3}}{2}, \pm \frac{1}{2})$, and $H_{0/z} (\bfG_{1,3,5}) = [H_{0/z} (\bfG_{2,4,6})]^* = C_{0/z} e^{i \phi_{0/z}}$. The potential terms are listed, along with the reciprocal lattice vectors, in Appendix \ref{app_moire}.
Equipped with this potential, the resulting {moir\'{e}}-reconstructed band structure can be computed for various hole dopings for the topologically trivial and non-trivial regimes.
The {moir\'{e}}-unit cell is given by $A_{\text{u.c.}} = \frac{\sqrt{3}}{2} a_m^2$. When there is one electron per spin per valley per unit cell, the {moir\'{e}} density is defined to be $n_s = 1/A_{\text{u.c.}}$.}
{Finally, a crucial complexity arises with implementing the \moire potential.
A finite ($V_m \neq 0$) periodic potential introduces band gaps at locations of band crossings (when $V_m = 0$), and in effect raises and lowers certain bands with respect to the chemical potential.
Thus, in order to maintain the electron occupation density in the {moir\'{e}} unit cell, it is necessary to permit the chemical potential to vary when the system is subjected to the {moir\'{e}} potential.
This shift of the chemical potential (in the presence of the \moire potential) has important ramifications with regard to the appearance of incompressible states at certain electronic densities.}
{In the following subsections, we examine the effects of the {moir\'{e}} potential over a range of broken symmetry phases occurring at electron densities (at $V_m=0$) between $[-2.65, 0] \times 10^{12} \text{cm}^{-2}$ for both $u_d >0$ (topologically non-trivial) and $u_d <0$ (topologically trivial) regimes.
We henceforth refer to the densities in terms of the hole filling per \moire unit cell, $\nu = n_e / n_s$, with $\nu = \pm 4$ referring to full filling.
It is important to keep in mind that if $\nu \in \mathbb{Z}$ (but not equal to $\pm 4$), though being an integer, nonetheless, is actually fractional filling of the \moire unit cell.
}

\subsection{Topologically Non-Trivial Regime}
\label{sec_moire_t_nt}

	\begin{figure}[h!]
	\center
		\includegraphics[width=0.85\linewidth]{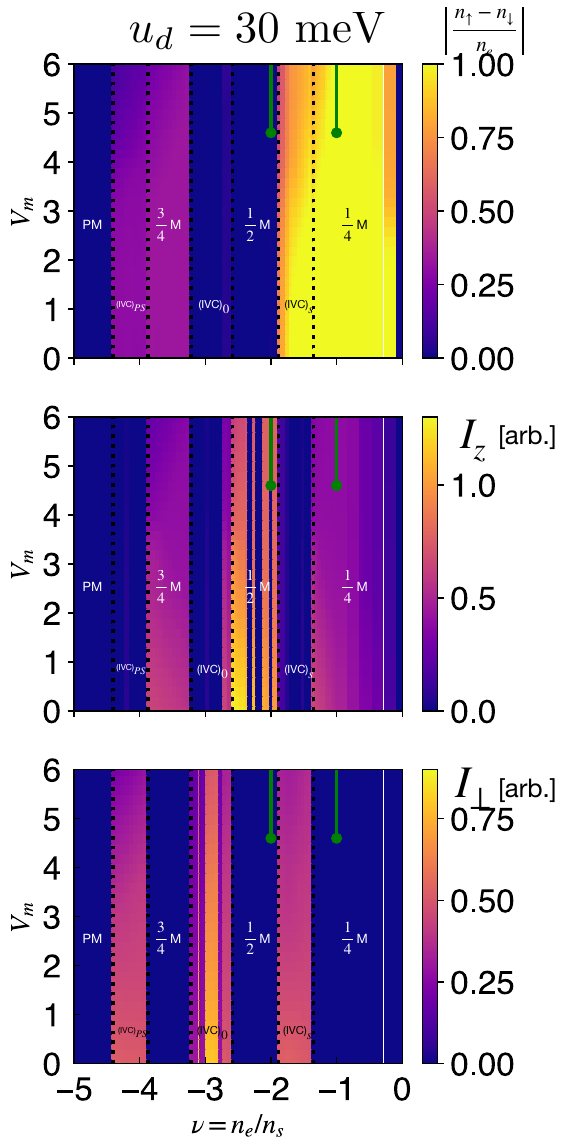} \\
		\includegraphics[width=0.48\linewidth]{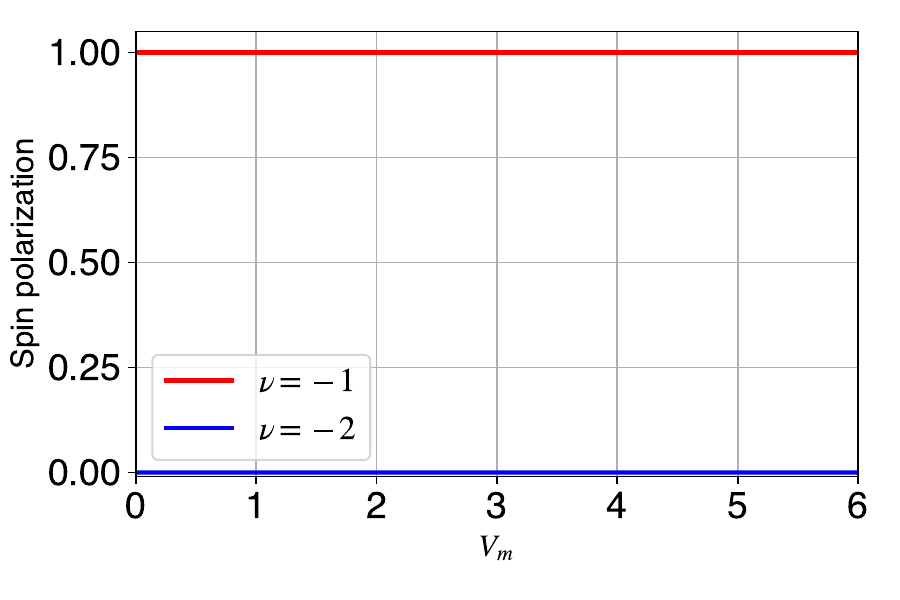} 
		\includegraphics[width=0.48\linewidth]{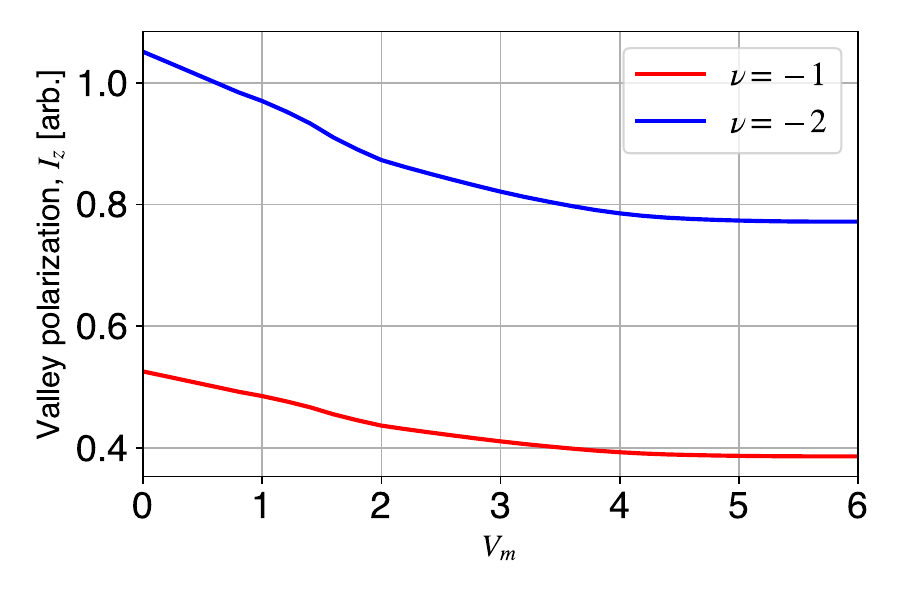}
		\caption{Top: Evolution of spin (top), valley polarization $I_z$ (middle), and inter-valley order $I_{\perp}$ (bottom) of the topologically non-trivial hole-doped Hartree-Fock phase diagram under the influence of the {moir\'{e}} potential ($V_m$). As seen the various order parameters are slightly weakened for increasing {moir\'{e}} potential strengths. The green lines indicate the appearance of the insulating states.
		The phases are denoted by their $V_m=0$ labels, with the fractional metals denoted as $\frac{m}{4}$M. $V_m = 1$ is the \textit{ab initio} estimate.
		Bottom: Line cuts for the spin and $I_z$ polarizations for $\nu = -1, -2$.}
    \label{fig_spin_polarized_evolution}

	\end{figure}

{We first examine the evolution of the $u_d>0$ HF-realized phases under the influence of the \moire potential}.
Figure \ref{fig_spin_polarized_evolution} depicts the spin and valley polarization, as well as the IVC order parameter, of the various phases as a function of {moir\'{e}} potential strength.
{As seen, for phases that begin with either partial or full spin polarization, an increasing {moir\'{e}} potential acts to only slightly weaken the net magnetization (and thus ferromagnetism) of a given phase.
Moreover, the IVC order and valley polarization of the phases is maintained as well (with slight changes), thus retaining the integrity of its {moir\'{e}}-less counterpart.
The only slight modification to the spin and valley character of the phases gives us confidence in our implementation of the {moir\'{e}} potential after accounting for the effects of Coulomb interactions i.e. in treating the {moir\'{e}} potential as a periodic perturbation to the HF bandstructure. 
It is this stability that enables us to confidently label the $V_m \neq 0$ phases under the label assigned to its corresponding $V_m=0$ phase.
We note that the $\frac{1}{2}$-metal phase in Fig. \ref{fig_spin_polarized_evolution} is not spin polarized, as we have selected one of the degenerate un-spin-polarized ansatzes.
}

	\begin{figure}[t]
	\center
		\includegraphics[width=0.99\linewidth]{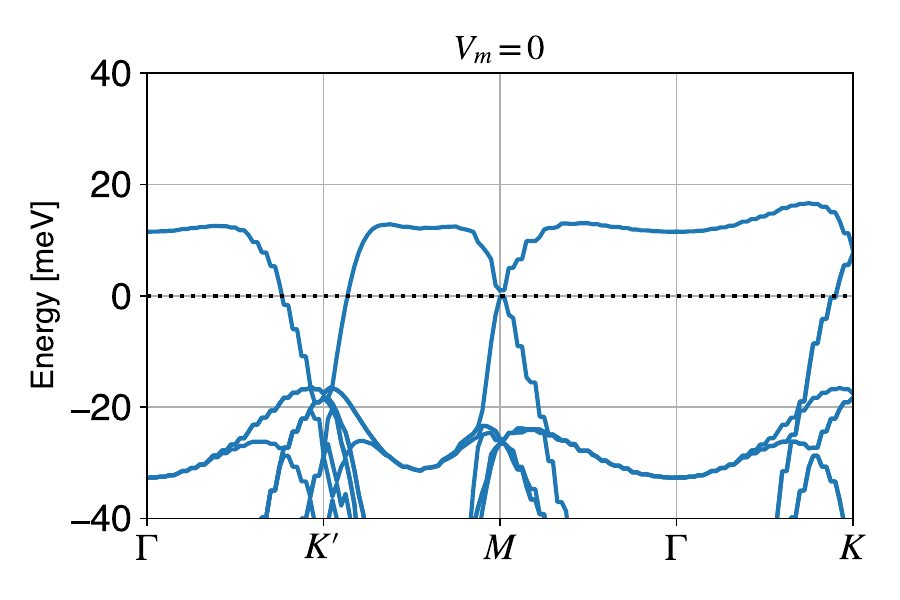} 
		\includegraphics[width=0.99\linewidth]{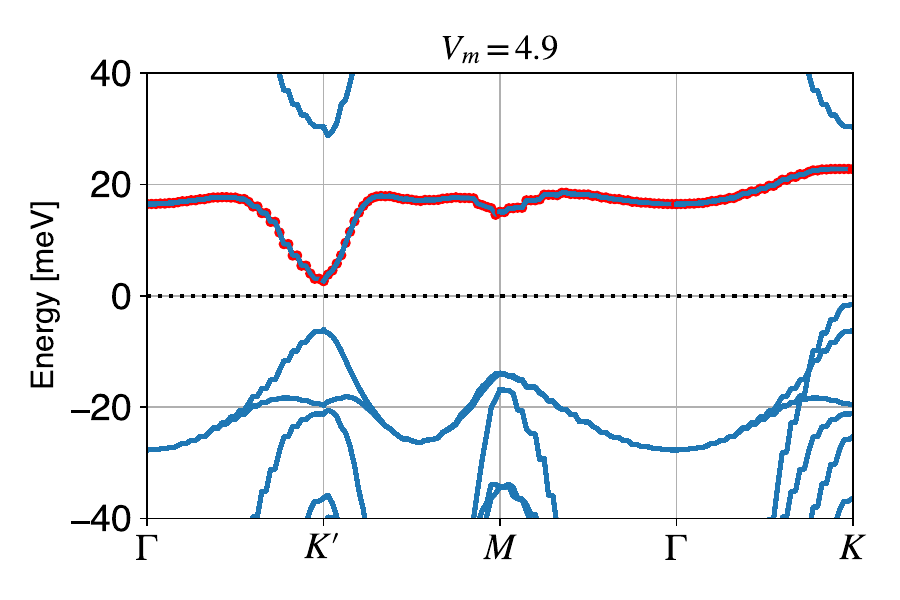}
		\caption{{Hartree-Fock band structure \textit{without} (\textit{with}) the {moir\'{e}} potential \textit{top} (\textit{bottom}) for $\nu = -1$ filling (i.e. $n_e = -n_s$) with $V_m=0$ and $V_m=4.9$, respectively, for the topologically non-trivial regime ($u_d = 30$ meV). The high-symmetry points are in the {moir\'{e}} mBZ (Fig. \ref{lst:python_code}). The dotted line represents the Fermi level/chemical potential; {for the $\nu = -1$ insulating band structure, the chemical potential can lie at any energy within the band gap.}
		The hole-valence band highlighted in red has a Chern number of $|C| = 3$.
		}
}
		\label{fig_band_structure_mbz}		
	\end{figure} 

{
The influence of the \moire potential is most apparent at fractional filling densities of the \moire unit cell, $\nu \in \mathbb{Z}$.
We present in Fig. \ref{fig_band_structure_mbz} the electronic bandstructure along high-symmetry directions in the mBZ for $\nu = -1$ ($n_e = -n_s$) for $u_d = 30$ meV with and without the \moire potential.
As seen in Fig. \ref{fig_band_structure_mbz}, the periodic {moir\'{e}} potential introduces band gaps, as well as the aforementioned raising and lowering of certain bands with respect to the chemical potential.
Evidently, incompressible (insulating) states appear at $\nu = -1$.}

The hole-valence band (highlighted in red) in Fig. \ref{fig_band_structure_mbz} is of a topological nature: it possess a Chern number of $|C|$ = 3, where the sign depends on whether the $K$ or $K'$ valley is being occupied by the parent (non-\moire) quarter-metallic phase.
We provide details on the computation of the Chern number in Appendix \ref{app_cn}.
\textcolor{black}{We note that the magnitude of the Chern number (calculated in this section in the sole presence of the screened Coulomb interaction) can be sensitive to the inclusion of interactions that allow the mixing of the active bands with the remote bands.
As discussed in Ref. \cite{feng_wang_abc_tlg}, a delicate mixing of active bands with remote bands has the possibility to reduce the Chern number.}

{Importantly as depicted in Fig. \ref{fig_spin_polarized_evolution}, correlated insulating gaps appear at densities centered at fillings $\nu = -1, -2$ with the gap increasing continuously as $V_m$ is increased beyond $\gtrsim 4.6$; we re-emphasize for clarity that $\nu = n_e / n_s$, and so insulators at $\nu = -1, -2$ are actually fractional filling of all of the spin-valley flavors.
Interestingly, these insulating states reside inside $\frac{m}{4}$-metallic phases, with the $\nu = -1 (-2)$ insulator appearing in the $\frac{1}{4}$ ($\frac{1}{2}$) metal phases.}

{
The appearance of insulating states arising from the $\frac{m}{4}$ metals is a striking feature for the $u_d>0$ regime.
Indeed, under the application of different displacement field strengths, the $\frac{m}{4}$ metallic phases and the IVC phases can be realized at different electronic densities.
As we depict in Appendix \ref{app_u_d_positive_other}, for larger (smaller) displacement fields, increasingly fewer (greater) number of $\frac{m}{4}$ metals are realized at integer $\nu$.
We thus discover that if $\nu = - m$ lies in the region of a $\frac{m}{4}$ metal (at $V_m = 0$), incompressible states are realized under the \moire potential ($V_m \neq 0$).
As a result (referring to Appendix \ref{app_u_d_positive_other} for the corresponding phase diagrams), for $u_d = 20$ meV, incompressible states are realized at all fractional fillings ($\nu = -1, -2, -3, -4$); while for $u_d = 25$ meV, they exist at $\nu = -1, -2, -3$; while for $u_d = 50$ meV, they exist at $\nu = -1$ solely
(for $u_d = 30$ meV shown in Fig. \ref{fig_spin_polarized_evolution} insulating states are only realized at $\nu = -1, -2$). {Evidently, increasing displacement fields can lead to the disappearance of band insulators ($\nu = -4$)}.
This relationship between the appearance of the insulator at integer-filling of the \moire unit cell and the corresponding spin-valley polarized $\frac{m}{4}$ metal at $V_m = 0$ highlights the importance of interaction effects on the electronic bandstructure.
 
}

{
The strength of the \moire potential required to realize the insulating states is also amplified as compared to the \textit{ab initio} studies estimates.
For $u_d \lesssim 30$ meV, the required $V_m$ above which insulators can be realized is $(V_m)_c \approx 4$, while for larger displacements $u_d = 50$ meV, $(V_m)_c \approx 7$.
{These insulating gaps open continuously at the aforementioned fillings above their respective critical \moire potential value $V_m$.}
The amplified {moir\'{e}} potential strength that is required to generate insulating states (as compared to the expectation from \textit{ab initio} studies estimates) once again highlights the substantial renormalization effects of bandstructure due to interaction effects.
{Indeed this amplification is not altogether surprising as mean-field studies typically underestimate the size of the energy gap (due to missing correlation effects); as such, \moire potential strengths larger than that estimated from \textit{ab initio} calculations are required to generate the observed insulator.
Such amplification thus indicates that strong (beyond mean-field) correlation effects are present for the topologically non-trivial regime.}
Moreover, the dependence of the (amplified from \textit{ab initio}) \moire potential strengths on the displacement field strengths suggests that accounting for interaction effects (in conjunction with varying displacement field strengths) may require going beyond the Hartree-Fock picture.
}

\subsection{Topologically Trivial Regime}

	\begin{figure}[h!]
	\center
		\includegraphics[width=0.77\linewidth]{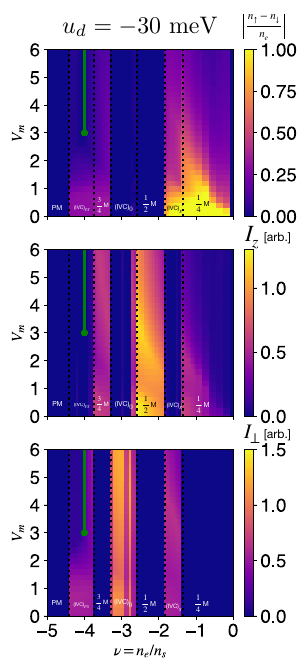} \\
		\includegraphics[width=0.47\linewidth]{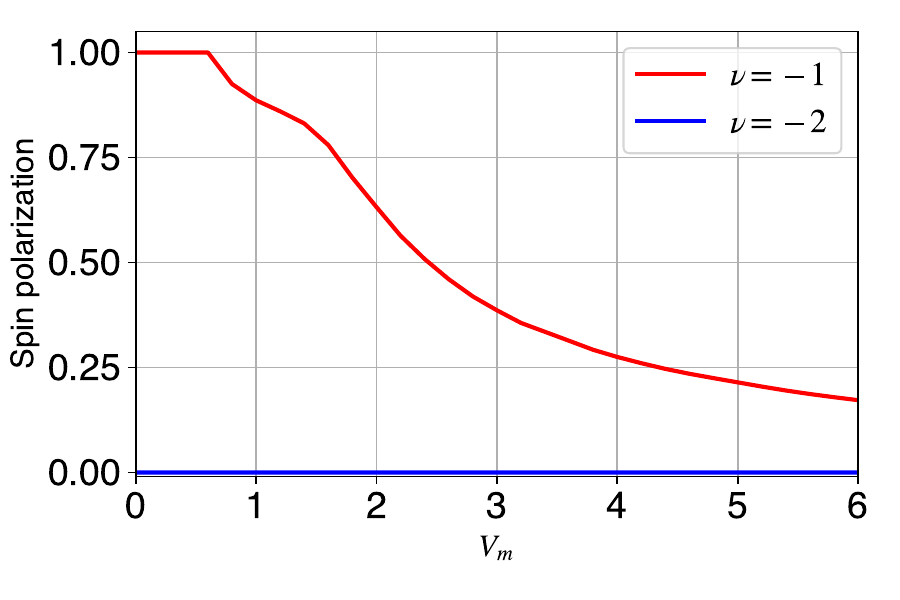} 
		\includegraphics[width=0.47\linewidth]{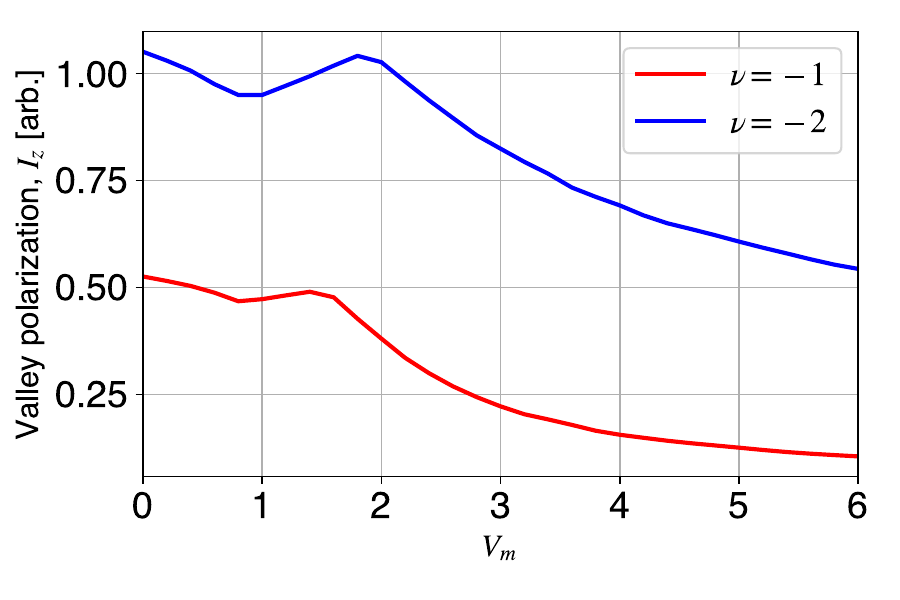}
		\caption{Top: Evolution of spin (top), valley polarization $I_z$ (middle), and inter-valley order $I_{\perp}$ (bottom) of the topologically trivial hole-doped Hartree-Fock phase diagram under the influence of the {moir\'{e}} potential ($V_m$). The dashed lines mark the location of the IVC order parameter ($I_x$) becoming finite. 
		As seen, the spin and valley integrity of the topologically trivial regime ($u_d <0$) undergo dramatic suppression by the {moir\'{e}} potential.
		{The delicacy of the phases leads us to denote labels for the $V_m \neq 0$ phases only near their $V_m = 0$ limit. $V_m = 1$ is the \textit{ab initio} estimate.
		Bottom: Line cuts for the spin and $I_z$ polarizations for $\nu = -1, -2$.}
}
		\label{fig_spin_polarized_evolution_n30}

	\end{figure}

	\begin{figure*}[t]
	\center
		\includegraphics[width=0.31\linewidth]{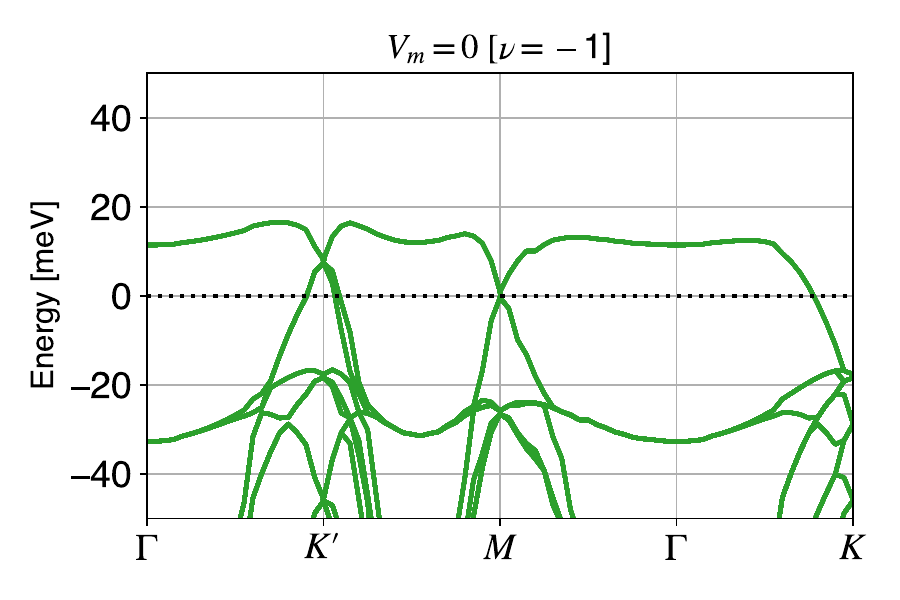} 
		\includegraphics[width=0.31\linewidth]{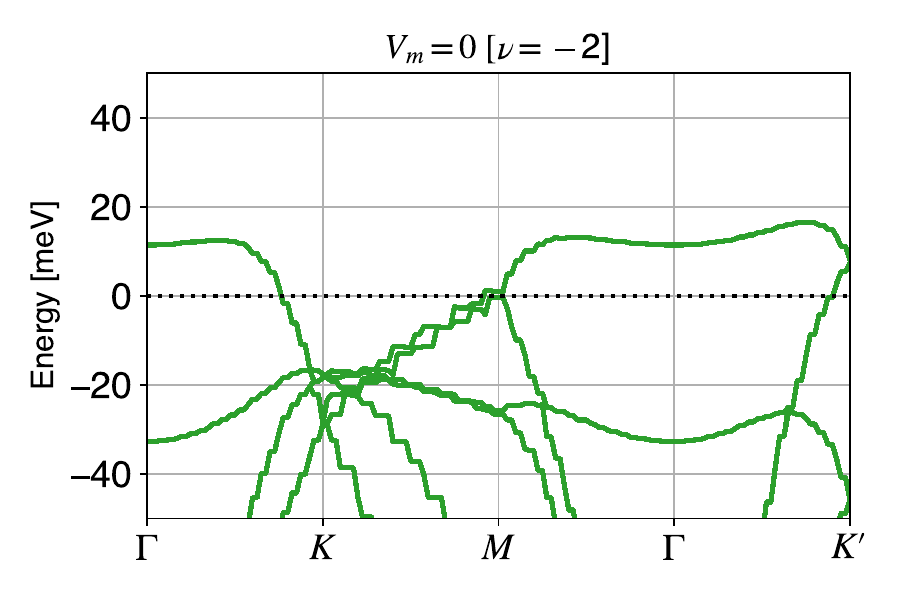}
		\includegraphics[width=0.31\linewidth]{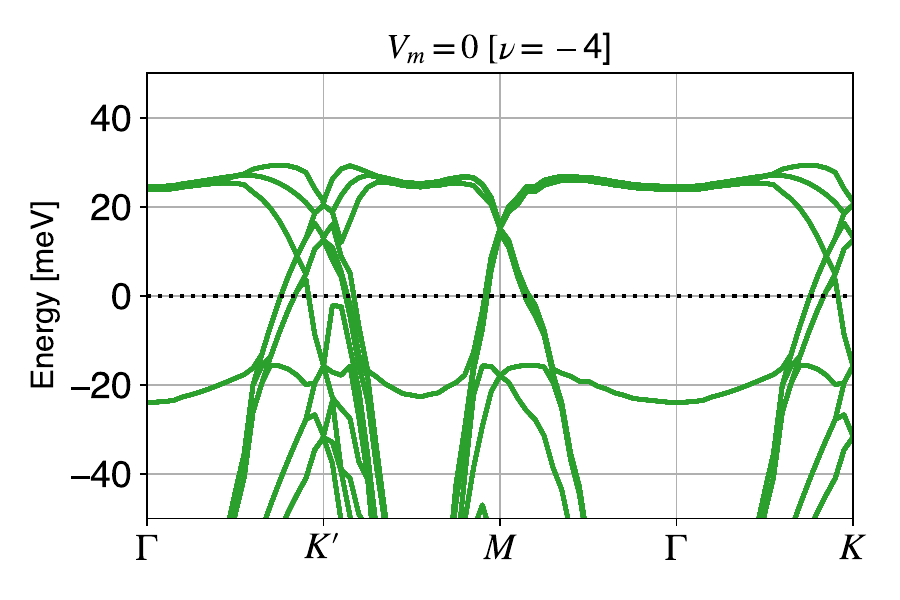}
		\\
		\includegraphics[width=0.31\linewidth]{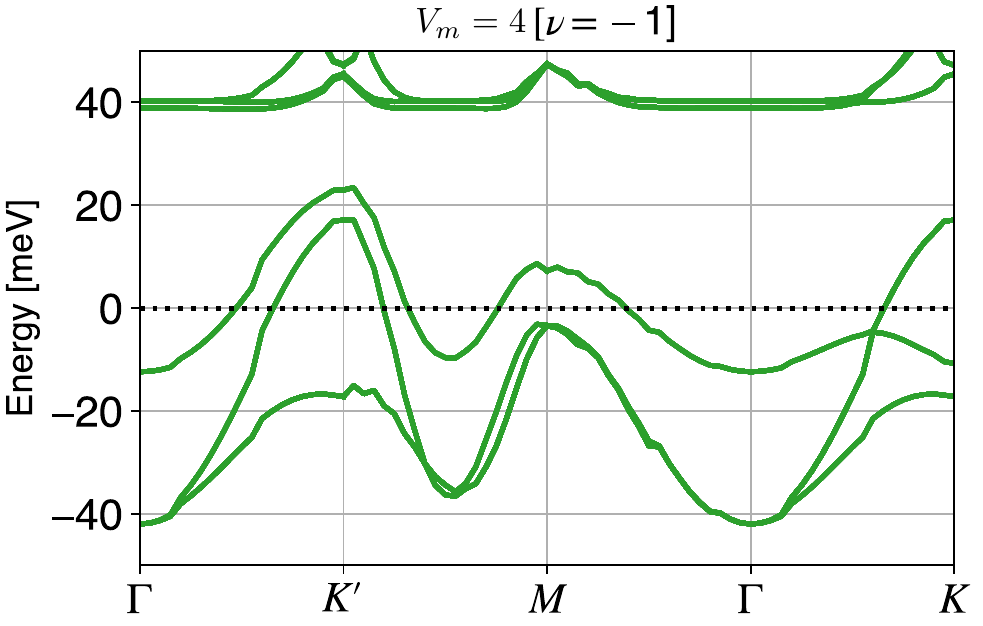}
		\includegraphics[width=0.31\linewidth]{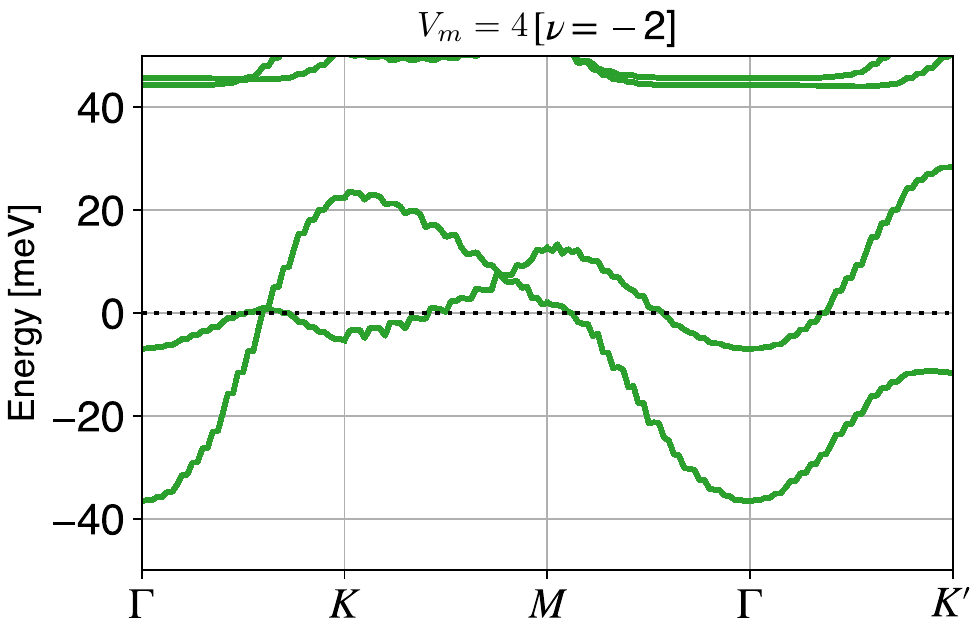}
		\includegraphics[width=0.31\linewidth]{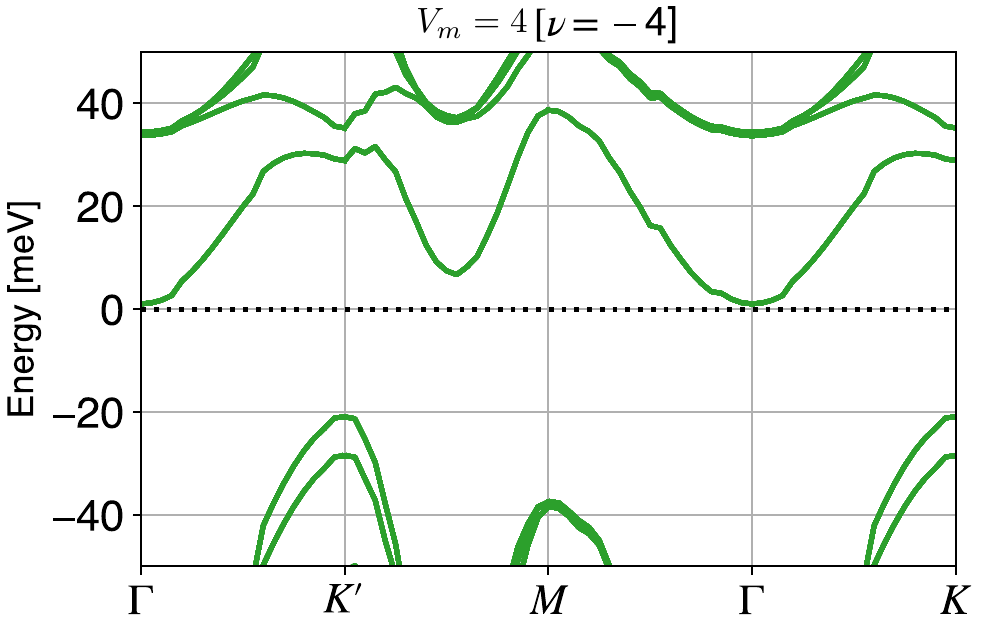}
		\caption{{Hartree-Fock band structure \textit{without} (\textit{with}) the {moir\'{e}} potential \textit{top} (\textit{bottom}) for $\nu = -1$ filling [left], $\nu = -2$ filling [center], $\nu = -4$ filling [right] with $V_m=0$ [top] and $V_m=4$ [bottom] for $u_d = -30$ meV. The high-symmetry points are in the {moir\'{e}} mBZ (Fig. \ref{lst:python_code}). The dotted line represents the Fermi level/chemical potential; {for the $\nu = -4$ band structure, the chemical potential can lie at any energy within the band gap.}
		}
}
		\label{fig_trivial_n1_band_structure_mbz}		
	\end{figure*} 
{The topologically trivial regime presents some surprisingly contrasting findings as compared to the $u_d >0$ regime.
We present in Fig. \ref{fig_spin_polarized_evolution_n30} the spin and valley polarizations, and the inter-valley order parameters (as defined in Appendix \ref{app_spin_order}, \ref{app_ivc_order}) under the application of the {moir\'{e}} potential.
As seen, the spin character in particular undergoes a dramatic suppression under the {moir\'{e}} potential; the IVC order parameter ($I_{\perp}$) undergoes a similar suppression in the $\frac{1}{4}$-metal phase.
Such changes highlight the delicate nature of the topologically trivial regime to increasing {moir\'{e}} potential strengths.}

{In conjunction with this delicacy is the occurrence of band insulating states solely at full-filling ($\nu = -4$) for $V_m \gtrsim 2-3$ (the precise value of the critical $V_m$ depends on the displacement field); this occurrence is also observed for varied displacement potential energies $u_d = -20$ meV and $u_d = -40$ meV as shown in Appendix \ref{app_u_d_neg_other}.
The appearance of an insulator at solely $\nu = -4$ (i.e. complete filling of all the four spin-valley flavors) though in agreement with band theory expectations, is not consistent with the aforementioned compressibility measurements where insulating peaks were also seen at $\nu = -1,2$.
{We present in Fig. \ref{fig_trivial_n1_band_structure_mbz} the HF bandstructure under the \moire potential for $V_M = 0$ and $V_m = 4$ for $\nu = -1, -2, -4$.
As seen, metallic behaviors for $\nu  = -1,-2$ are retained even with an amplified \moire potential, while an energy gap has developed for $\nu= -4$.}
Such findings suggest that the treatment of the {moir\'{e}} potential as a weak perturbation on top of the Hartree-Fock renormalization effects is not an appropriate description for the topologically trivial regime.
Indeed, it highlights that a strong-coupling analysis may be required to accurately capture the effects of correlations along with the} {moir\'{e} potential in this regime \cite{briding_hubbard_senthil}.
}

\subsection{Comparison of topologically non-trivial and trivial regimes}

{
The evidently distinct behaviors for the $u_d>0$ and the $u_d<0$ regime can be traced back to the role of the applied displacement field.
The displacement field acts in a staggered manner on the two active layers: in the kinetic energy term, Eq. \ref{eq_h0}, it appears as $\sim u_d \sigma^3$ where we recall that the Pauli matrix is in the ($A_1$, $B_3$) layer basis.
For $u_d>0$ ($u_d<0$) such a staggered potential lowers the $B_3$ ($A_1$) layer state in energy with respect to the $A_1$ ($B_3$) layer state.
This relative lowering of energy between the layers leads to an enhanced occupation of one layer over the other for the various broken-symmetry phases; we recall that the Coulombic interaction in Eq. \ref{eq_screened_coulomb} is energetically impartial to either of the two layers, and so the layer partiality is solely from the kinetic energy (there is, of course, some mixing between the layers from the remaining terms in Eq. \ref{eq_h0} as well as from the interaction Eq. \ref{eq_screened_coulomb}).
}

{
The impact of this asymmetry in the $A_1$ and $B_3$ layers is brought to the forefront in the \moire setting, where h-BN is aligned with the top layer of ABC-TLG and consequently only applies a potential to the $A_1$ layer.
With $u_d > 0$, the hole occupation is predominantly on the $B_3$ layer, and so an $A_1$-layer-acting potential
does not significantly alter the nature $B_3$-layer flavor polarized states.
As such, the \moire potential has a relatively limited influence and only weakly perturbs the nature of the HF-states, which is in agreement with the only slight modification of the spin and valley characters of the phases in Fig. \ref{fig_spin_polarized_evolution} and Appendix \ref{app_u_d_positive_other}.
On the other hand, with $u_d<0$ the hole occupation is predominantly in the $A_1$ layer, and so an $A_1$-layer-acting potential
can have a significant impact on the nature of eigenstates (as seen in Fig. \ref{fig_spin_polarized_evolution_n30} and Appendix \ref{app_u_d_neg_other}). 
}

\section{Discussion}
\label{sec_discussion}

{In this work, we quantitatively examined the role of an underlying h-BN substrate on ABC-stacked TLG under the influence of varied displacement field strengths. Our primary goal was to examine the suggestion in Ref. \cite{tlg_abc_young} that the \moire potential induced by the h-BN alignment serves to produce a band gap at the chemical potential of the correlated metal that exists in moir\'{e}-less ABC trilayer graphene. Our main conclusion is that such a picture may be tenable in the topologically non-trivial regime but does not seem to be so in the topologically trivial regime. 

Employing a Hartree-Fock mean-field analysis, and focusing on the hole-doped regime, we showed that for the topologically non-trivial regime (i) prominent incompressible states appear at certain fillings, $\nu \in \mathbb{Z}$, of the {moir\'{e}} unit cell, and (ii) the magnetization of the spin-polarized phases is slightly weakened under ever-increasing {moir\'{e}} potential strengths.
The incompressible states appear whenever $\frac{m}{4}$-metallic states coincide with integer $\nu = -m$ fillings of the \moire unit cell (an occurrence facilitated by changing the displacement field strengths).}

{We compare our findings with the experimental identification of insulating states at $\nu = -1, -2$ for moderate displacement fields, while for small displacement fields there is also an occurrence of insulating states at solely full-filling ($\nu = -4$) \cite{tlg_abc_young}.
Firstly, we address the observation of the $\nu  = -3$ insulator in our mean-field studies.
Since the $\frac{3}{4}$-metal -- the originator of the $\nu = -3$ insulator -- has been shown to be suppressed (on the electron doped regime \cite{tlg_abc_young}) via the inclusion of an inter-valley Hund's coupling, we are similarly inclined to believe the corresponding $\nu = -3$ insulator will be suppressed.
Indeed, incorporating the effects of inter-valley Hund's coupling systematically with accurate estimates of the Hund's coupling (there have been promising results on this front from electron spin resonance in twisted bi-layer graphene \cite{tblg_hund_esr}) would be an interesting direction to consider in future works.
Secondly, the observed broad trend of disappearing insulators at increasing displacement field strengths is in general agreement with our findings, where we notice that the $\nu = -4$ band insulator is the first to be suppressed by the increasing displacement field strengths.
Why, however, an amplified \moire potential field strength (as compared to \textit{ab initio} estimates) is required to observe the insulators in both the $u_d > 0$ and $u_d<0$ regime, and whether this is an artifact of the Hartree-Fock analysis is another open question that warrants further investigations. 
}

{
For the topologically trivial regime, we found sharp insulating peaks appearing solely for complete filling of the \moire unit cell ($\nu = -4$) which occurs in conjunction with the strong suppression of the {moir\'{e}}-less ferromagnetism. 
Similar to the non-trivial regime, an amplified \moire potential (as compared to \textit{ab initio} estimates) is required to develop the insulator.
This apparent difference in the impact of the \moire potential on the topologically trivial and non-trivial regimes is in contrast to the insulating peaks appearing in transport \cite{feng_wang_abc_tlg} and compressibility measurements \cite{tlg_abc_young} for both positive and negative displacement fields \cite{tlg_abc_young}. 
More specifically, for the topologically trivial regime, the lack -- in the presented theory -- of correlated insulating states at $\nu = -1,-2$ suggests that {describing the system in terms of weak-coupling itinerant flavor-polarized phases whose bands are reconstructed by the \moire potential is not an appropriate description.
Indeed, the findings from this work suggest that a strong-coupling formalism may be necessary to capture the effects of the interactions in the presence of a \moire potential.}}
{
The recent spectroscopic study \cite{long_ju_abc_tlg} in ABC/h-BN provides corroborating indications of the highly correlated nature of the $u_d<0$ regime.
Optical transitions, facilitated by irradiating the sample with an infrared bream, provide a measure of the energy gap (from the peak in the photocurrent spectrum), as well as the single-particle bandwidth from the observed peak width.
Aligning h-BN with the top layer, and examining the photocurrent spectra for both positive and negative displacement fields ($D$), 
 it was observed that the peak width was slightly narrower for $D<0$, as compared to $D>0$. 
The suppressed bandwidth for $D<0$ demonstrates that correlation effects are stronger and thus play a more dominant role for the $u_d<0$ regime.
This indication of strong correlation effects for $u_d <0$, as well as the observed asymmetry between the positive and negative displacement field directions, is in agreement with our findings.

Thus, a more appropriate starting point for the topologically trivial regime will be a lattice model that is obtained by projecting the Coulomb interaction to the active \moire bands. Such a model was presented in Ref. \cite{briding_hubbard_senthil}. Thus in this case the \moire potential is clearly important in determining the nature of the correlated states. 
Even in the topologically non-trivial regime, the amplification of the \moire potential required to produce the correlated insulator suggests that perhaps \moire is important even in that situation. 
}

\acknowledgements
We thank Zhiyu Dong, Shubhayu Chatterjee, Perry T. Mahon, and Geremia Massarelli for helpful discussions. TS was supported by NSF grant DMR-2206305, and partially through a Simons Investigator Award from the Simons Foundation to Senthil Todadri. This work was also partly supported by the Simons Collaboration on Ultra-Quantum Matter, which is a grant from the Simons Foundation (651440, TS)
The authors acknowledge the MIT SuperCloud and Lincoln Laboratory Supercomputing Center for providing HPC resources that have contributed to the research results reported within this manuscript.
\break

\appendix

\section{ABC-TLG Parameters}
\label{app_params}

The tunneling parameters in Eq. \ref{eq_h0} {follow those proposed in Ref. \cite{tlg_abc_young}}: $\gamma_0 = 3.1$ eV, $\gamma_1 = 0.38$ eV, $\gamma_2 = -0.15$ eV, $\gamma_3 = -0.29$ eV, $\gamma_4 = -0.141$ eV, $\delta = -0.0105 eV$, $u_a = -6.9 \times 10^{-3}$ eV, and $v_\alpha = \frac{\sqrt{3}}{2} \gamma_\alpha$, where $\alpha \in [0,4]$.
{As discussed in Ref. \cite{tlg_abc_young}, these parameters were obtained by comparing ABC-stacked to ABA-stacked (Bernal) trilayer graphene, where penetration field capacitance measurements enabled estimates of certain (common to ABC) hopping parameters \cite{aba_tight_binding_estimates}.}
The displacement field is accounted for by $|u_d| = 30$ meV in the main text figures, and the interaction potential parameters in Eq. \ref{eq_screened_coulomb} are $d_s = 150 a_0$ and $\epsilon = 8$.

For the Hartree-Fock studies without the {moir\'{e}} potential we employ a UV cutoff of $k_{max} = 0.1 a_0^{-1}$, with convergence testing performed upto $N=91 \times 91$ mesh grid for the electron doped regime.
When including the {moir\'{e}} potential, due to the additionally required momentum points in $\bfk + \bfG$ in Eq. \ref{eq_moire_potential}, we employ a UV cutoff of $k_{max} = 0.2 a_0^{-1}$ with $N=131 \times 131$ mesh grid ({with convergence consistency established up to $N=141 \times 141$ for $u_d = 30$ meV and up to $N=141 \times 141$ for $u_d = -20$ meV}) for the Hartree-Fock studies for the hole-doped regime.
The Hartree-Fock mean field self-consistent solutions are obtained using standard fixed point iteration schemes (in conjunction with Steffensen's methods to boost computation speed) {with moderate parallel computing resources (up to $\sim40$ cores for \moire potential implementation)}.
\textcolor{black}{The self-consistent iteration (at fixed electron density) is performed by starting with an initial ansatz that is both spin and valley conserving (to obtain the $\frac{m}{4}$-metals), only spin conserving (to allow for realizing IVC phases), and one that is zero at all momentum points (this case generates on the first iteration mean field expectation values consistent with the non-interacting model).
We take a random (with different random seed) collection of initial expectation values at the different momentum points for each of the ansatzes.
Upon the completion of the self-consistent loop, the average energy is evaluated and compared to those from the other ansatzes, with the lowest energy solution selected.
}

{The exact location of the phase boundaries (in density) for the various displacement field strengths can be determined more precisely with finer mesh sizes and with more number of electronic density points; nonetheless, the types of broken-symmetry phases, their locations, and the development of the insulating phases, are sufficiently determined with the aforedescribed mesh size and number of electronic densities.
For clarity, the phase diagrams were generated with eighty-three density points (with the exception of $u_d = 25$ meV, where one-hundred-and-thirty-five density points were used to carefully discern the phase boundary close to the $\nu = -3$ insulator).
}

\section{Implementation of {Moir\'{e}} Potential}
\label{app_moire}

The {moir\'{e}} potential reconstructs the original Hartree-Fock bandstructure into a mini-Brillouin zone (mBZ). 
We follow the standard scheme to implement the periodic potential, namely for each momentum in the mBZ we construct a large matrix spanned by the translation by the reciprocal lattice vectors $\bfg_j = m \bfg_+  + n \bfg_-$ \cite{nearly_flat_bands_senthil}.
{
We recall that the HF band structure is implemented in a UV box with an appropriately discretized momentum mesh; we label each point in the UV box as $\bfq_i$.
In order to implement the \moire potential over the mBZ, we discretize the mBZ, and for each point $\bfk_i$ we find the momentum point ($\bfq_i$) in the original UV box which is closest (in momentum space distance) to it.
This procedure is repeated for successive $\bfk + \bfg_j$, thus enabling the HF solution to be implemented with the \moire potential efficiently. 
}

\begin{figure}[h]
\centering
\begin{minipage}{0.5\linewidth}
\label{fig_moire_bz_vec}
\subcaption{}
\includegraphics[scale=0.5]{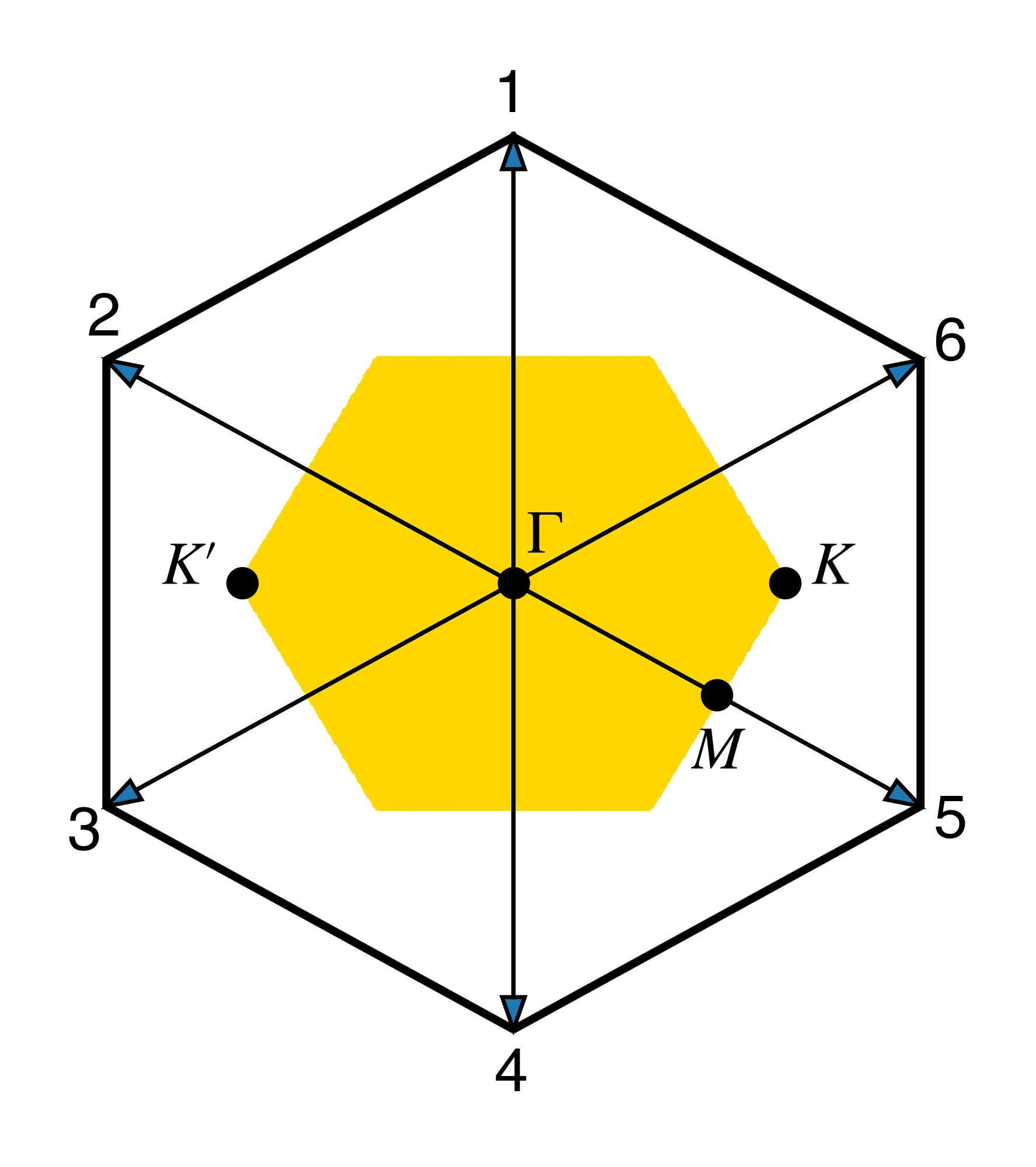}
\end{minipage}\hfill
\begin{minipage}{0.4\linewidth}
\subcaption{}
\begin{tabular}{c|c}
$\bfG_j$ & $(m,n)$ \\ \hline
$\bfG_1$ & $(1,-1)$ \\
$\bfG_2$ & $(1,0)$ \\
$\bfG_3$ & $(0,1)$ \\
$\bfG_4$ & $(-1,1)$ \\
$\bfG_5$ & $(-1,0)$ \\
$\bfG_6$ & $(0,-1)$ \\
\end{tabular}
\end{minipage}
\caption{{Moir\'{e}} reciprocal lattice vectors imposed on the mini Brillouin zone (mBZ) in gold with labelled high-symmetry points. Location of reciprocal lattice vectors (a). Each solid black line contributes to the {moir\'{e}}-potential in Eq. \ref{eq_moire_potential}. The arrow heads indicate the specific location of the reciprocal lattice vector $\bfg_j = m \bfg_+  + n \bfg_-$ in the mBZ (b), where $\bfg_{\pm} = \frac{4 \pi}{\sqrt{3} a_m} (-\frac{\sqrt{3}}{2}, \pm \frac{1}{2})$.}
\label{lst:python_code}
\end{figure}

The reciprocal lattice vectors are defined in Fig. \ref{lst:python_code} are $\bfg_j = m \bfg_+  + n \bfg_-$ and $\bfg_{\pm} = \frac{4 \pi}{\sqrt{3} a_m} (-\frac{\sqrt{3}}{2}, \pm \frac{1}{2})$. 
In our present context, we implement seven such bases to generate the complete {moir\'{e}}-influenced bandstructure.
The {moir\'{e}} potential parameters as listed in \cite{moire_band_abc_tlg_theory_macdonald, nearly_flat_bands_senthil} are $C_0 = -10.13$ meV, $C_z = -9.01$ meV, and $\phi_0 = 86.53 ^\circ$, $\phi_z = 8.43 ^{\circ}$.
We note that we have (for the sake of demonstration) examined the impact of the relaxation effects of a graphene layer on hBN for $u_d = \pm30$ meV.
In particular, we considered the impact of local stacking modifications due to strains (in Ref. \cite{PhysRevB.96.085442} this is referred to as case of ``Relaxed $\alpha$''), and found that the phase boundaries and the conclusions of our work (based on the above ``rigid'' parameters) do not get modified.

\section{Spin Order Parameter: Magnetization}
\label{app_spin_order}

{
Due to the lack of appreciable spin-orbit coupling, the mean field solutions are diagonal in spin space.
Thus the spin polarization of a phase can be defined by considering the spin-up and spin-down sectors of the (spin-diagonal) mean field Hamiltonian separately. 
We define the eigenvalues for the spin-up (spin-down) sectors to be $\xi^{\uparrow (\downarrow)}_{, \bfk}$.
With these defined spin eigenvalues, the density of spin-up and down electrons (measured from with respect to CNP) are,
\begin{align}
n_{s} = \frac{1}{A} \sum_{\bfk} \left[\sum_{m} n_F(\xi^{s}_{\bfk}) - N^s_\text{occupied} \right],
\label{eq_spin_order}
\end{align}
where $m$ sums over all the band and internal (valley, layer) flavors, $n_F(...)$ is the Fermi-Dirac function, and we subtract out the filled bands via $N^s_\text{occupied}$; in the presence of the {moir\'{e}} potential, $N^s_\text{occupied} = 14$, which is half (i.e. occupied number at CNP) of the spin-up bands.
As a brief remainder, we recall that the total number of bands $7 \times 2 \times 2 \times 2$, where the 7 is from the number of {moir\'{e}} reciprocal lattice vectors (as shown in Fig. \ref{lst:python_code}) and each of the factors of 2 correspond to the spin, valley, and layer, respectively.
}
\section{IVC order parameters: Valley Polarization and IVC order}
\label{app_ivc_order}

{	
The valley polarization and the degree to which the valley degrees of freedom can ``mix'' provides helpful measures to characterize a phase.
In particular, one can define valley order parameters that indicate (i) the magnitude of the mixing between ($I_{x,y}$), and (ii) the magnitude of the valley polarization ($I_z$), for each electron density $n_e$,
\begin{align}
I_a = \left| \sum_{\alpha, \beta, s} \sum_{\Lambda, \Omega} \frac{1}{A} \sum_{\bfq} \Big[ \langle c^{\dag}_{\Lambda, s, \alpha} (\bfq) c_{\Omega, s, \beta} (\bfq)  \rangle \tau^a _{\Lambda, \Omega} \Big] \right|,
\end{align}
where $a = \{x,y,z\}$, $\tau^a$ specifies a Pauli matrix in valley space (with $\Lambda, \Omega$ indicating the valley labels $\bfK$, $\bfK'$), $\{\alpha, \beta\}$ are over the band degrees of freedom (including the layer label), and $s$ being the spin quantum number.
The spin $s$ is diagonal due to the lack of appreciable spin-orbit coupling, as described in Appendix \ref{app_spin_order}.
Since both $I_{x,y}$ indicate valley mixing, we define a perpendicular component to indicate the overall IVC mixing, $I_{\perp} = \sqrt{I_x^2 + I_y^2}$.
\textcolor{black}{We include a unit of area normalization of $[10^{16} m^2]^{-1}$ for the valley and IVC order parameters.}}

\section{Hartree-Fock results without the \moire potential} 

\label{app_hf_no_moire}

Since our primary interest in this work is in the hole-doped regime, we briefly discuss the features of the electron doped regime, before turning to the hole-doped regime

\subsection{Electron-doped symmetry-broken phases}

{The $\frac{1}{4}$-metal breaks time-reversal symmetry, and is spin-polarized (with finite magnetization as defined in Appendix \ref{app_spin_order}) as well as valley polarized (as defined in Appendix \ref{app_ivc_order}) which entails an appreciable anomalous Hall effect in transport measurements.}

The $\frac{1}{2}$-metallic phase is found to occupy any of the two-flavors: same-valley opposite-spin, opposite-valley same-spin, and opposite-valley and opposite-spin (same-valley same-spin is forbidden by Pauli principle).
This variety of possible phases for the $\frac{1}{2}$-metal, as well as the existence of the subsequent $\frac{3}{4}$-metal, is a manifestation of the {earlier discussed time-reversal symmetry and valley-independent spin rotation symmetry;}
{as was examined in Ref. \cite{tlg_abc_young}, lowering the symmetry, by the inclusion of Hund's coupling -- of magnitude $|J_H| > 0.03 V^{\text{sc}}_C (0)$ for the inter-valley Hund's coupling -- removes the $\frac{3}{4}$-metal region of the phase diagram.}
Finally, at high dopings, the kinetic energy prevails over the Coulombic interaction to result in a democratically populated (in both spin and valley flavors) unpolarized paramagnetic metal (PM).

As the electron doping density is increased, within each $\frac{m}{4}$-metallic phase, the occupation of the $m$-flavors are identical and continue to increase until entering the next $m+1$ phase (where another flavor is occupied) upon which the electron density rearranges so as to equally populate all of the $m+1$ flavors ({all of which} continue to grow until this phenomenon repeats at the next instability).
Such {democratic} occupation of the electron density amongst the various flavors is in contrast to the `cascading' picture in TBG, where (as the electron density is increased) upon encountering each Stoner instability, one of the flavors reaches maximum occupation while the remaining electron densities are `reset' to empty \cite{cascade_tbg_ilani_pablo, cascade_tbg_yazdani}.

\subsection{Hole-doped symmetry-broken phases}

Increasing hole doping from CNP, we first encounter the familiar $\frac{1}{4}$-metal, which is the familiar time-reversal broken phase realized in the electron-doped phase.
{In this phase, the Fermi surface is composed of a single band and the corresponding wavefunctions possess non-vanishing amplitude for only the single flavor being occupied.
As a result of the single flavor occupation, the Fermi surface geometry breaks $C_6$ symmetry.
These properties are reflected in Fig. \ref{fig_spin_polarized_evolution} by the pronounced spin polarization and valley polarization $I_z$.}

One next encounters a fully spin-polarized IVC phase, (IVC)$_{S}$; more specifically, one of the spin species is completely occupied while the other spin species is empty. 
Indeed, the occupation of one degenerate spin-flavor over the other is spontaneously chosen.
{For an aforementioned ferromagnetic IVC-Hund's coupling, this spin polarized IVC state will be weakly favored.}
The nature of this phase is reflected in the band at the Fermi level having wavefunction contributions from both valleys i.e. each momentum location on the Fermi surface is composed of a superposition (in same spin species) of both $K$ and $K'$ valleys, as is expected for an IVC phase.
{This occupation of both valleys is reflected in a vanishing valley polarization $I_z$ and the finite IVC order $I_{\perp}$ as shown in Fig. \ref{fig_spin_polarized_evolution}.}
As hole doping penetrates deeper into (IVC)$_S$ and close to the phase boundary to the $\frac{1}{2}$-metal, the occupation of the Fermi surface spreads across multiple bands, accompanied by a slight weakening of the spin-polarization; the occupation of additional bands at the Fermi level accounts for the lack of pure spin-polarization close to the transition to the $\frac{1}{2}$-metal {(this slight weakening of the spin polarization is seen for other $u_d>0$ displacement fields as shown in Appendix \ref{app_u_d_positive_other}).}

{The familiar $\frac{1}{2}$-metallic phase arrives next. 
In this phase, the Fermi surface is made up of two bands, with each band carrying one of the orbital/flavor characters (that are being occupied in this phase).
Depending on which flavors are occupied, the $C_6$ symmetry can be preserved or broken (for example, for the opposite-valley same-spin half-metal, $C_6$ is preserved over bands).
This phase can possess a finite valley polarization $I_z \neq 0$ (as defined in Appendix \ref{app_ivc_order}) if both valleys are occupied; this possibility is clearly depicted in Fig. \ref{fig_spin_polarized_evolution}.}
{With the inclusion of the aforementioned antiferromagnetic/ferromagnetic Hund's coupling, the opposite-valley, opposite/same-spin state is energetically selected.}

Following this $\frac{1}{2}$-metal phase, one encounters spin-unpolarized ({weakly favored by an antiferromagnetic IVC-Hund's coupling}) IVC metal, (IVC)$_{0}$, where both spin species are equally populated (we contrast this with fully-spin polarized IVC phase, (IVC)$_S$, that was encountered at lower hole dopings).
For the densities upon just entering the phase, the flavor occupation at the Fermi level is spread across two bands, and the deeper one penetrates into the phase, multiple bands (at each momentum location) possess a linear combination of the valley characters.
{Such properties are once again reflected in the finite $I_{\perp}$ in Fig. \ref{fig_spin_polarized_evolution}.}

Following this (IVC)$_{0}$, one next encounters a $\frac{3}{4}$-metallic phase.
{For the $\frac{3}{4}$-metal, the Fermi surface is spread over three bands, with each band occupying one of three flavors.}
This is followed by a partially spin-polarized IVC metal, (IVC)$_{PS}$, wherein both spin species are occupied, however with an imbalance in their spin leading to their spontaneous (partial) magnetization.
In the (IVC)$_{PS}$, one of the bands at the Fermi surface possess an IVC character (i.e. linear combination of both valleys), while the other two bands possess a single flavor occupation.
{We note that the $\frac{3}{4}$-metal has total normalized spin-magnetization of $\frac{1}{3}$ (due to three spin-valley flavors being occupied) with the (IVC)$_{PS}$ being slightly reduced from this (in darker shades of purple in the colormap in Fig. \ref{fig_spin_polarized_evolution}).}
{Indeed, these properties of the partially spin polarized IVC phase are depicted by the prominent (yet not fully) polarized spin as well as a non-negligible $I_{\perp}$ in Fig. \ref{fig_spin_polarized_evolution}}.

{Finally, upon further increasing the density, one encounters the spin and valley unpolarized paramagnetic metal $(PM)$, reflected in the vanishing valley, IVC and spin polarization order parameters in Fig. \ref{fig_spin_polarized_evolution}, with the $C_6$-symmetric Fermi surface spread over four bands (with each band occupying one of the four spin-valley flavors).}
We note that upon increasing the density into the hole-doped regime a variety of possible broken-symmetry phases may be realized (depending on the displacement field, strength of interactions etc.) \cite{zalatel_abc_tlg, macdonald_tlg_2022}.

\section{Impact of {moir\'{e}} potential at displacement field strengths for $u_d >0$}
\label{app_u_d_positive_other}

\begin{figure*}[t!]
	\center
		\includegraphics[width=0.31\linewidth]{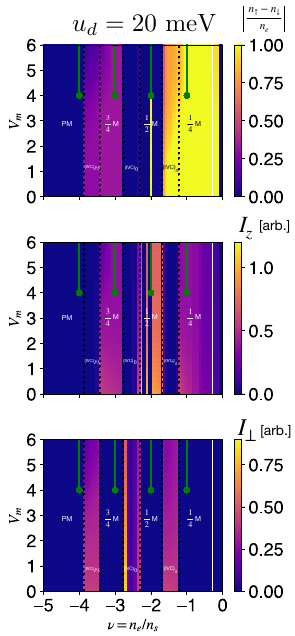}
		\includegraphics[width=0.33\linewidth]{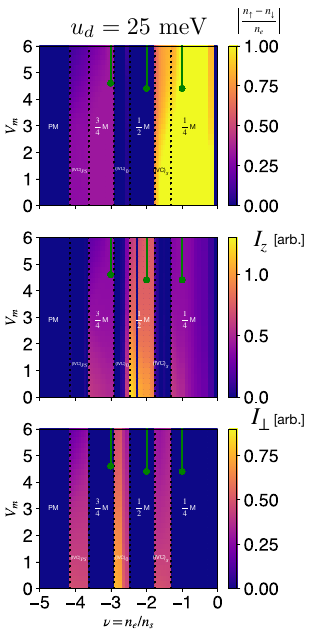}
		\includegraphics[width=0.32\linewidth]{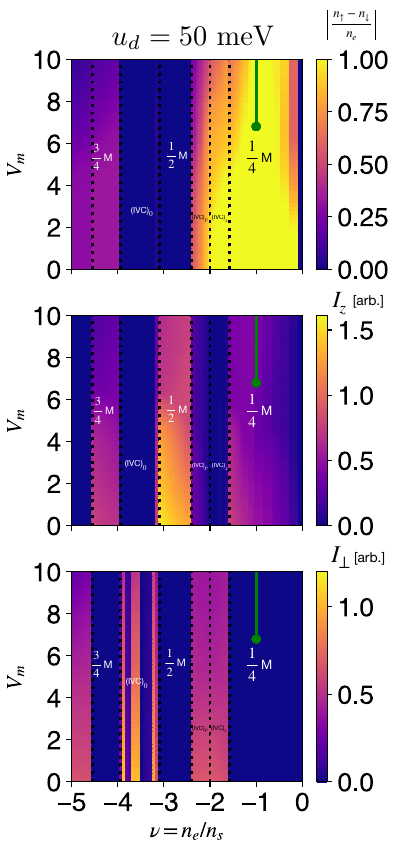}
		\caption{ {Evolution of spin (top), valley polarization $I_z$ (middle), and inter-valley order $I_{\perp}$ (bottom) of the topologically non-trivial hole-doped Hartree-Fock phase diagram under the influence of the {moir\'{e}} potential ($V_m$) for $u_d = 20$ meV [left], $u_d = 25$ meV [center], and $u_d = 50$ meV [right]. Just as with $u_d = 30$ meV, the spin polarization (of initially spin-polarized phases) is slightly weakened for increasing {moir\'{e}} potential strengths. $V_m = 1$ is the \textit{ab initio} estimate.
		Unlike $u_d = 30$meV, incompressible states (indicated by the green lines) appear at all integers $\nu$ for $u_d = 20$ meV for $V_m \gtrsim 4.0$; at $\nu = -1,-2,-3$ for $u_d = 25$ meV for $V_m \gtrsim 4.4$; and at $\nu = -1$ for $u_d = 50$ meV for $V_m \gtrsim 6.8$. The dashed lines mark the location of the IVC order parameter ($I_x$) becoming finite; for $u_d = 50$ meV, it also marks the location where the spin polarization weakens in the (IVC)$_S$. 
		 The (IVC)$_P$ phase is characterized as having inter-valley coherent order, as well as partial spin polarization and partial valley polarization (hence the subscript ``P'').}}
		\label{fig_spin_polarized_evolution_25}
	\end{figure*}

{We present in Fig. \ref{fig_spin_polarized_evolution_25} the corresponding phase diagram for $u_d = 20, 25, 50$ meV (topologically non-trivial) under an ever increasing {moir\'{e}} potential.
The general nature of the $V_m=0$ phases for the $u_d = 30$ meV carry over; we thus use the same labelling scheme for the phases as in the main text.
However, the crucial difference, as discussed in the main text, is the additional/reduced number of insulating states realized under the \moire potential depending on the strength of the displacement field.
For small displacement fields ($u_d = 20$ meV), insulators occur at $\nu = -1, -2, -3, -4$; for $u_d = 25$ meV, insulators occur at $\nu = -1, -2, -3$; for large displacement field insulators only occur at $\nu  = -1$.}

\section{Impact of {moir\'{e}} potential at displacement field strengths for $u_d <0$}
\label{app_u_d_neg_other}
	
\begin{figure*}[t!]
	\center
		\includegraphics[width=0.48\linewidth]{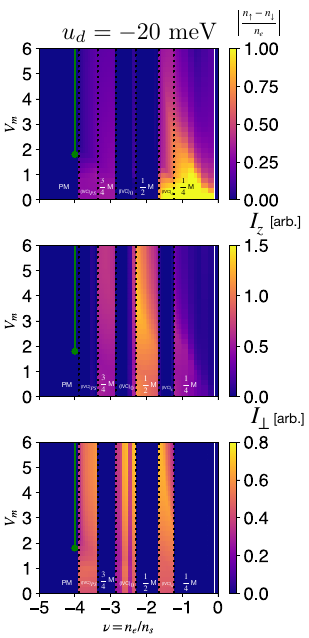}
		\includegraphics[width=0.47\linewidth]{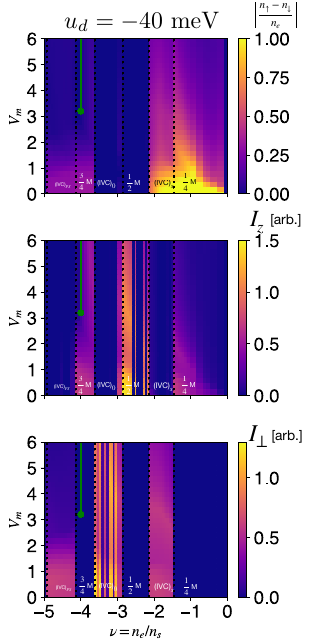}
		\caption{{Evolution of spin (top), valley polarization $I_z$ (middle), and inter-valley order $I_{\perp}$ (bottom) of the topologically trivial hole-doped Hartree-Fock phase diagram under the influence of the {moir\'{e}} potential ($V_m$).
		Left: $u_d = -20$ meV. Right: $u_d = -40$ meV. The dashed lines mark the location of the IVC order parameter ($I_x$) becoming finite. 
		As seen, the spin and valley integrity of the topologically trivial regime ($u_d <0$) undergo dramatic suppression by the {moir\'{e}} potential. $V_m = 1$ is the \textit{ab initio} estimate.
		Incompressible states appear (approximately $2-3$ times amplified potential strengths as compared to \textit{ab initio} estimates) solely at $\nu = -4$, complete filling, as expected from band theory. }
		{The delicacy of the phases leads us to denote labels for the $V_m \neq 0$ phases only near their $V_m = 0$ limit.}}
	\label{u_d_m20_m40}
	\end{figure*}

{Figure \ref{u_d_m20_m40} depicts the impact of the {moir\'{e}} potential for $u_d = -20$ meV and $u_d = -40$ meV (i.e. topologically trivial regime).
Similar to $u_d = -30$ meV, the HF phases are more delicate than their topologically non-trivial counterparts to the periodic potential.
Unlike the topologically non-trivial regime, we obtain the solitary insulator at $\nu = -4$ over a wide range of displacement fields.}
	
{ 
\section{Chern number for $\nu=-1$ band}
\label{app_cn}
In the non-interacting model, for $u_d>0$, there is a well isolated band which allows a well-defined Chern number to be defined for the valence band, $C = \pm 3$, where the sign depends on which valley is being occupied \cite{moire_band_abc_tlg_theory, nearly_flat_bands_senthil}.
Due to time-reversal symmetry, the Chern number associated with each valley is related to each other by a minus sign.

In the hierarchy of energy scales considered in this work, for the $u_d>0$ regime, there is a well-isolated hole-valence band in Fig. \ref{fig_band_structure_mbz} for a range of amplified $V_m \in [4.6, 5.4]$; for $V_m \gtrsim 5.4$, the band collides with an above remote band.
As such, we focus on $V_m = 4.9$ for the computation of the Chern number for $u_d = 30$meV.
We compute the Chern number following Ref. \cite{chern_method_2005,moire_band_abc_tlg_theory}, where the Berry curvature is efficiently computed using a gauge-invariant Wilson loop.
For the sake of self-containedness we briefly sketch out the rationale behind this methodology.

For each momentum point $\bfk$ in the mBZ, we extract out the Bloch-eigenvector associated with the hole-valence band (in this present work, this is the 27$^{\text{th}}$ band), $\ket{\psi(\bfk)}$.
In order to define the Wilson loop, we consider the amplitude defined by,
\begin{align}
    U_{i} (\bfk) = \frac{\braket{\psi(\bfk + \delta \hat{i})|\psi(\bfk)}}{|\braket{\psi(\bfk + \delta \hat{i})|\psi(\bfk)}|},
\end{align}
where $\hat{i} = \{\hat{x}, \hat{y}\}$ are unit vectors in the mBZ grid, and $\delta$ is the distance between two points in the square-discretized grid.
We note that the normalized amplitude is a complex phase.
For each point point $\bfk$, we can then simply define a Wilson loop by the product of such phase-amplitudes around a square plaquette,
\begin{align}
    W(\bfk) = U_{\hat{x}}(\bfk)U_{\hat{y}}(\bfk + \delta \hat{x}) [U_{\hat{x}}(\bfk + \delta \hat{y})]^* [U_{\hat{y}}(\bfk)]^*. 
\end{align}
Since, under a gauge transformation, the Bloch state transforms as $\ket{\psi(\bfk)} \rightarrow \ket{\psi(\bfk)} e^{i\lambda(\bfk)}$, one can explicitly check that the above product is invariant under a gauge transformation.
The gauge-invariant Wilson loop is related to the gauge-invariant Berry curvature ($\Omega(\bfk)$), 
\begin{align}
    \Omega(\bfk) = \frac{\Log[{W(\bfk)}]}{\delta^2}, 
\end{align}
where we use the Principal logarithm.
Thus, the Chern number can be computed from the integral of the Berry curvature in the mBZ, 
\begin{align}
    C = \frac{1}{2 \pi i} \sum_{\bfk \in \text{mBZ}} \delta^2 \Omega(\bfk) 
\end{align}
where the momentum integral is already discretized.
}

\break 
\break


%

\end{document}